\documentclass[reqno,]{tatra13a_sciendo_CC_AOP}         

\usepackage{hyperref} 
\makeatletter \thm@headfont{\bfseries\scshape} \makeatother
\usepackage[labelfont=bf,labelsep=period]{caption}

\usepackage{latexsym,euscript}
\usepackage{amsbsy,amscd,amsfonts,amsmath,amsopn,amssymb,amstext,amsthm,amsxtra,cases}
\usepackage{epsf,epsfig, graphics, color, verbatim}
\usepackage{cite, float, multirow}
\usepackage[font=small]{caption}
\usepackage{caption}
\usepackage{subcaption} 
\theoremstyle{plain}
\theoremstyle{plain}

\theoremstyle{definition}

\def \R {{\mathbb{R}}}

\issueinfo{}{}
\pagespan{1}{36}
\begin{document}

\title
[Segmentation of the Aorta from CT Data] {Mathematical and numerical methods for accurate aorta segmentation from non-enhanced CT Data yielding reliable identification and evaluation of large vessel vasculitis}

\author[1,2]{Konan A. Allaly}    
\affil[1]{Department of Mathematics and Descriptive Geometry, Slovak University of Technology, Bratislava, SLOVAKIA}
\address[1]{
            Department of Mathematics and Descriptive Geometry\\
            Faculty  of Civil Engineering \\
            Slovak University of Technology\\
            Radlinskeho 11, 81005 Bratislava \\
            SLOVAKIA
           }
\email{ allaly.anderson@stuba.sk}
\email{ karol.mikula@stuba.sk }
\author[2]{Jozef Urb\'an}
\address[2]{
            TatraMed Software s.r.o.\\
            L\'i\v{s}\v{c}ie \'udolie 9, 84104 Bratislava\\
            SLOVAKIA
           }
\email{ konan.allaly@tatramed.sk }
\email{ jozef.urban@tatramed.sk }
\author[1]{Karol Mikula}
\affil{TatraMed Software s.r.o., Bratislava, SLOVAKIA}

\def\shortauthors{KONAN A. ALLALY --- JOZEF URB\'AN --- KAROL MIKULA}

\keywords{Segmentation, minimal path, curve evolution, large-vessel vasculitis}
\thanks {This work was supported by grants Marie Sklodowska-Curie No. 955576, APVV-23-0186, and VEGA 1/0249/24.}



\begin{abstract}
Segmentation of the aorta is crucial for various medical analyses, such as the diagnosis and treatment of cardiovascular diseases. This work presents mathematical models and methods yielding a semi-automatic segmentation of the aorta from non-enhanced CT data. Our framework consists of three steps. First, using the minimal path approach, we extract a path within the aorta from two user-supplied points. Then, using 3D Lagrangian curve evolution, we move the initial path to the approximate centerline of the aorta. The centered path is used in the last step to construct the initial condition for the generalized subjective surface method (GSUBSURF). Applying the GSUBSURF method with this initial condition yields an accurate segmentation of the aorta. The segmentation results and the manual segmentations overlap, with a worst-case mean Hausdorff distance of $2.175 \pm 0.605$ mm for a voxel spacing of $0.977$ mm. Using the aorta centerline and segmentation, we define precise regions of interest along the aorta to assess large-vessel vasculitis from patient FDG-PET/CT image data. The application shows promising results, as we demonstrated widespread inflammation throughout the aorta in a patient before treatment. After treatment, we observed a significant reduction in inflammation while accurately identifying the aorta regions where inflammation persisted. These findings also align with those of experienced medical doctors who have worked on the same cases.

\end{abstract}

\maketitle

\section{Introduction}\label{Introduction}

Vasculitis is the inflammation of blood vessels of various sizes. Early detection and treatment are crucial to prevent irreversible damage, such as loss of vision in retinal vasculitis and aortic dissection when the aorta is involved \cite{Castellanos_2013}. Various approaches are used to diagnose vasculitis, including biopsy and imaging. Numerous studies conducted by rheumatologists, radiologists, clinicians, and researchers \cite{Castellanos_2013, Arnaud_2009, Stenova_2009, Slart_2018} have demonstrated the role of $^{18}$F-fluorodeoxyglucose-positron emission tomography (FDG-PET) imaging in the classification, diagnosis, and follow-up of vasculitis. Two approaches are used to interpret FDG-PET images: visual and quantitative analysis \cite{Slart_2018}. A radiologist performs the visual analysis, and the result depends on the user's experience. For the quantitative analysis, regions of interest (ROIs), including the vessel wall, are manually extracted and used for the interpretation. Due to the low resolution of PET images, and to better define the ROIs, FDG-PET images are often combined with another scan in the same coordinate system, such as Computed Tomography (CT), where anatomical structures are more clearly visible. Therefore, this work focuses on providing an accurate segmentation of the aorta from CT data to define precise ROIs.

\bigskip

\begin{figure}[ht]
\begin{center}
\includegraphics[scale=0.6]{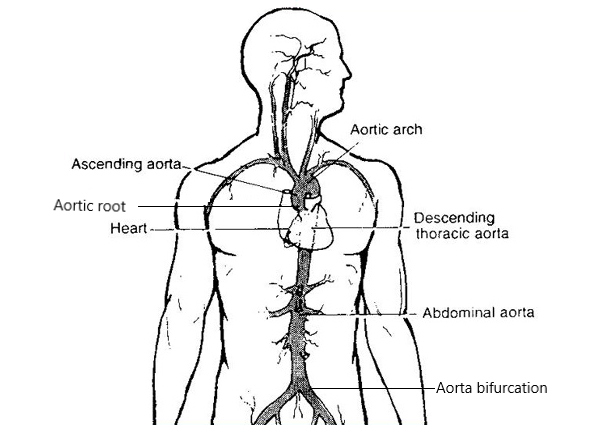}
\end{center}
\caption{Aorta location in the body \cite{Kovacs_2010}} 
\label{Aorta_location} 
\vspace{-5pt}
\end{figure}

The aorta is the main artery of the human body, carrying oxygen-rich blood from the heart to the rest of the body. The aorta is a tube-like structure divided into anatomical segments, including the aortic root, ascending aorta, aortic arch, descending aorta, and abdominal aorta. The aorta's section located in the thorax (aortic root, ascending aorta, aortic arch, and descending aorta) is called the thoracic aorta, and the section in the abdomen is called the abdominal aorta (see Fig.\ref{Aorta_location}). 

Due to its importance, researchers have been working intensively on segmenting the aorta from magnetic resonance imaging (MRI), computed tomography angiography (CTA), and CT images \cite{Kovacs_2010, Kurugol_2012, Gamechi_2019, Xie_2014, Trullo_2017, Yuan_2021}. Segmenting the aorta from non-enhanced CT data is challenging due to noise, weak or missing edges, low contrast, and closeness to other organs with similar intensity. Two main approaches are used to detect the aorta in non-enhanced CT data. Some algorithms use a priori models built from manually segmented training data. The other algorithms use anatomical information and the Hough Transform to detect the aorta in a given slice as a circular shape. Thresholding and region growing \cite{Adams_1994, Bronzino_2000} are widely used to segment medical images, yielding promising results when the image contrast is high. However, they lead to over-segmentation for low-contrast or highly noisy images. Deformable models \cite{Kass_1988, Caselles_1997_b, Sarti_2000} are also used for medical image segmentation, but they require a good initial condition and strong edges to achieve optimal results. Recently, machine learning and deep learning algorithms have been investigated and used to segment the aorta \cite{Yuan_2021, Trullo_2017}. Various approaches have been proposed to segment the aorta across different image modalities. The available works proposed for non-enhanced CT data focused only on the thoracic aorta. Kurugol et al. \cite{Kurugol_2012} proposed an automatic segmentation of the aorta using Hough Transform and level set in non-contrast chest CT data. The algorithm detects the ascending and descending aorta by Hough transform in the heart region. The centers of detected circles are used to reconstruct the aorta's initial surface. The initial surface is finally refined with 3D level set evolution. Lower segmentation accuracy is observed around the aortic arch, which is attributed to the limitations of the ground-truth accuracy. Martinez-Mera et al. \cite{Martinez_2013} proposed segmenting the thoracic aorta in contrast-enhanced CT angiography (CTA) images using a combination of level-set and region-growing algorithms. They use the Hough transform to locate the aorta and define region-growing seed points in the original image. The regions (aortic arch, descending aorta, and ascending aorta) were grown using the mean and variance of the neighboring voxels. The level set method is then used for refinement and segmenting the missing parts. Performance decreased in the heart region, where boundaries with the contrast-filled cardiac chambers are blurred. Gamechi et al. \cite{Gamechi_2019} developed a fully automatic segmentation of the thoracic aorta from non-contrast-enhanced CT data. Multi-atlas registration is used to localize the aorta, then seed points are set in the ascending and descending aorta as the center of mass of the initial aorta surface estimation. The aortic centerline is then extracted from the defined seed points. An optimal surface graph cut technique, initialized through a dilated centerline, is employed to extract the aortic surface. Trullo et al. \cite{Trullo_2017} proposed a deep learning framework for joint segmentation of thoracic organs in CT images, specifically the aorta, heart, esophagus, and trachea. The approach used a SharpMask-based fully convolutional network (FCN) to combine low-level and high-level features for improved localization. The approach requires a large amount of labeled data for the training phase, and that is often not freely available in medical image processing. 

\bigskip

This work presents a robust framework to segment the entire aorta (thoracic aorta and abdominal aorta) from non-enhanced CT data. Our framework consists of three steps. First, we extract, from two given points, a path within the aorta using the minimal path approach \cite{Cohen_1997}. Usually, the minimal path approach is applied to images with high contrast in the region of interest. However, we define a new potential function that combines image intensity and edge detection, making it applicable to our low-contrast input images. In the second step, we utilize a 3D Lagrangian curve evolution model \cite{Mikula_Urban_2014} to move the initial path to the approximate centerline of the aorta. In the Lagrangian curve evolution, the motion of a given curve is defined by a vector field. In \cite{Mikula_Urban_2014}, the vector field is determined from the segmented colon. However, we do not have the segmented aorta. Therefore, we define a new vector field from the edge detector. In the third step, we utilize the centered path to construct the initial condition for the GSUBSURF method \cite{Mikula_2009, Mikula_Peyrieras_2007, Mikula_Peyrieras_2011}, which ultimately yields the final segmentation. The GUSBSURF provides satisfactory results when the initial segmentation is appropriate. For example, to segment a circular structure, a good initial condition is constructed from a point at the center of mass of the structure \cite{Sarti_2000}. Therefore, we use the aorta centerline to build the initial condition. The contribution of this paper is thus to provide a complete framework to extract the entire aorta (thoracic and abdominal aorta) from non-enhanced CT data. Each step of our framework is robust enough to handle the variability of aorta shape and can segment both healthy and unhealthy aortas from CT image data with or without contrast-enhanced material.

\bigskip

The paper is organized as follows: in Section 2, we present the extraction of a path within the aorta, and in Section 3, we present the centering of the extracted path. In Section 4, we present the GSUBSURF model and its application to aorta segmentation, discussing our segmentation results. In Section 5, we present the application of our segmentation framework to the diagnosis of large vessel vasculitis. In the final section, we conclude our work and emphasize the global implications of our study. 

\newpage
\section{Path extraction within the aorta}

In this section, we extract a path within the aorta using the minimal path approach. The minimal path approach was introduced in \cite{Cohen_1997} as a global minimization formulation of the snake model \cite{Kass_1988} using the \textit{Fast Marching} method \cite{Sethian_1996}. It simplifies the initialization of the snake model by setting just two endpoints. The authors start by converting the snake energy into a path integral. Then, they define the minimal action map and show that its level-set propagation implies the Eikonal equation. They finally solved the Eikonal equation by the \textit{Fast Marching} method and back-traced the solution to obtain the globally minimal path. In \cite{Cohen_1997}, the model was applied to road extraction in 2D aerial images and to organ segmentation in 2D medical images. The model has been extended to 3D in \cite{Deschamps_2001} for applications such as virtual endoscopy in medical images. Computational improvements, such as partial propagation starting from one endpoint and stopping at the second one, were introduced to make the model effective for real-time applications. In \cite{Benmansour_2009}, the authors introduced another improvement, in which a single point is specified and new points are automatically detected during the front propagation. Each newly detected point is set as a new starting point to create a recursive growing path. This improvement allows the extraction of a path in an elongated 2D/3D structure. The authors \cite{Cohen_1997, Deschamps_2001, Li_2007, Benmansour_2009} mentioned that the quality of the extracted path depends strongly on the definition of the potential function. In noisy and complex 3D structures, such as the aorta in non-enhanced CT data, paths may cross the vessel wall, leading to invalid paths. To overcome this issue, we introduce a potential function that allows computation of a path within the aorta in a single step. In cases with poor contrast, a growing approach is used to ensure the path remains within the aorta.

\bigskip
\subsection{Minimal path approach}

Let us consider a parametric curve $\Gamma(v) : \Omega \to \mathbb{R}^{3}~$, $v \mapsto (x(v), y(v), z(v))$ in a 3D image, where $\Omega = [0,1]$ is the parametrization interval and $v$ a parameter. For a space $\mathcal{A}$ of admissible paths $\Gamma$, the snake energy $E: \mathcal{A} \to \mathbb{R}$ has the following form
\begin{equation}
    \Gamma \mapsto E(\Gamma) = \int_{\Omega} \left ( ~\frac{\omega_1}{2} ||\Gamma'(v)||^2 + \frac{\omega_2}{2} ||\Gamma''(v)||^2 + P(\Gamma(v)) ~\right ) dv \label{snake_energy}
\end{equation}
where $\Gamma'$, $\Gamma''$ are the first and second derivatives of $\Gamma$ with respect to $v$, $||\cdot||$ is the Euclidean norm in $\mathbb{R}^{3}$ , $P$ is the potential function defined from the image, $\omega_1$ and $\omega_2$ are the model parameters \cite{Kass_1988, Cohen_1997}. Contrary to the classical snake energy \cite{Kass_1988}, the authors of \cite{Cohen_1997}, chose $v$ as the arc-length parameter $s$, which means $||\Gamma'(s)|| = 1$, then $\Omega = [0,L]$, where $L$ is the length of the curve. This makes the energy depend only on the geometric curve and not on the parametrization \cite{Cohen_1997}. The second derivative of $\Gamma$  is removed from the energy (\ref{snake_energy}). These changes yield a model in which initialization involves setting only two endpoints and the potential function responsible for curve smoothing. The energy of the new model $~E : \mathcal{A}_{p_s, p_f} \to \mathbb{R}~$ has the following form
\begin{equation}
    \begin{aligned}
        \Gamma \mapsto E(\Gamma) &= \int_{0}^{L} \left ( ~\omega ||\Gamma'(s)||^2 + P(\Gamma(s)) ~\right )ds \\
        &= \int_{0}^{L} \left( \omega + P(\Gamma(s)) \right) ~ds  = \int_{0}^{L} \Tilde{P}(\Gamma(s))~ds  \label{snake_energy_minimal_path}
    \end{aligned}
\end{equation}
where $~\Tilde{P} = \omega + P~$ is defined from the image, $\mathcal{A}_{p_s, p_f}$ is the space of all curves connecting  the starting point $p_s$ and a final point $p_f$. The boundary conditions are given by $\Gamma(0)=p_s$, and $\Gamma(L) = p_f$. The parameter $\omega > 0$ is used to control the smoothness of the curve.

At each point $p$ in the image domain, and for $\Tilde{P} > 0$, the minimal action $U_{p_s}$ corresponds to the minimal energy integrated along a path that starts at $p_s$ and ends at $p$
\begin{equation}
    U_{p_s}(p) = \inf_{\Gamma(L) = p} \left( \int_{\Gamma} \Tilde{P} ds \right) = \inf_{\mathcal{A}_{p_s, p}} E(\Gamma) \label{inf_action}
\end{equation}
$\mathcal{A}_{p_s, p}$ is the set of all paths between $p_s$ and $p$.

To compute the minimal action $U_{p_s}$, the authors of \cite{Cohen_1997} formulated the following partial differential evolution equation (related to (\ref{inf_action}))
\begin{equation}
    \frac{\partial \mathcal{L}(s,t)}{\partial t} = \frac{1}{\Tilde{P}} \vec{n}(s,t) \label{pde_minimal_path}
\end{equation}
for the set of equal energy isosurface $\mathcal{L}$ in "time"
\begin{equation}
    \mathcal{L}(s,t) = \{p \in \R^{3} ~|~ U_{p_s}(p) = t\}
\end{equation}
where $\vec{n}(s,t)$ is the normal to $\mathcal{L}(\cdot, t)$. The equation (\ref{pde_minimal_path}) is a front propagation equation with a speed $~\tfrac{1}{\Tilde{P}}~$ where $t$ represents the time at which the front passes over the point $p$ in the image domain. It starts with an initial surface represented by a small sphere centered at the starting point $p_s$ until $U_{p_s}$ is computed for all points in the image domain. During the front propagation, the head of the front moves faster in regions with smaller potential values $\Tilde{P}$. We denote the head of the front as the point farthest (in terms of the local norm) from the starting point reached by the front. In fact, the authors of \cite{Cohen_1997} also showed that $U_{p_s}$ satisfies the Eikonal equation
\begin{equation}
    ||\nabla U_{p_s} || = \Tilde{P}, ~~~~~  U_{p_s}(p_s) = 0.
\end{equation}
In this work, similarly to \cite{Cohen_1997}, we employ the \textit{Fast Marching} method \cite{Sethian_1996} to compute $U_{p_s}$, as it is fast and has lower computational complexity compared to other approaches \cite{Cohen_1997, Deschamps_2001}.

The gradient of $U_{p_s}$ is orthogonal to the propagating front. Therefore, the minimal path $\Gamma_{p_s,p_f}$ between any point $p_f$ and the starting point $p_s$ can be found by back-propagation on the map $U_{p_s}$. To find $\Gamma_{p_s,p_f}$, the authors in \cite{Cohen_1997, Deschamps_2001} suggest the \textit{Steepest Descent Gradient} method
\begin{equation}
    p_{i+1} = p_i - \tau~\frac{\nabla U_{p_s}}{||\nabla U_{p_s}||}, ~~ \tau > 0, ~~ i = 0,...,M-1  \label{DG}
\end{equation}
where $\tau$ is taken as $0 < \tau < 1$ and $M$ is the maximal number of iterations. The algorithm starts with $p_0 = p_f$ and stops when the Euclidean distance between the current point $p_i$ and $p_s$ is less than a given tolerance. 

\bigskip

\subsection{Potential function}\label{define_potential}
The potential function $\Tilde{P}$ governs the evolution of the front in the \textit{Fast Marching} front propagation. The definition of the potential function depends on the application. An effective potential exhibits strong contrast between the regions of interest, their boundaries, and the surrounding background. In practice, low potential values are assigned to pixels or voxels in the target region, whereas high values are assigned to background voxels \cite{Cohen_1997, Deschamps_2001, Benmansour_2009}. With such a function $\Tilde{P}$, we expect the front to reach regions of interest before spreading to less relevant regions, leading to the extraction of the expected path. However, erroneous paths may be extracted when the potential is noisy, insufficiently contrasted, or when the target structure is elongated, thin, or highly curved \cite{Benmansour_2009}. In segmentation tasks, for instance, the potential function is typically designed to take smaller values along object boundaries \cite{Cohen_1997}. In our application, in contrast, smaller values are assigned to voxels along the aorta centerline and larger values to voxels along the aorta walls, so that the boundaries act as barriers confining the front within the aorta. In the following sections, $I$ denotes the input image filtered by the geodesic mean curvature flow filtering \cite{Kriva_2010}. For our application, we propose the following potential function
\begin{equation}
\Tilde{P}(p) = \left( \omega + \frac{|\Tilde{I}(p_s) - I(p)|}{\underset{q \in \mathcal{I}}{max}(|\Tilde{I}(p_s) - I(q)|)} \right) \frac{1}{1 + d(p)}, ~\forall p \in \mathcal{I}, ~p \neq p_s  \label{potential}
\end{equation}
where $\mathcal{I} \subset \mathbb{R}^d$, $d = 2,3$, is the image domain, $\Tilde{I}(p_s)$ is the average image intensity value in a spherical volume centered at the starting point $p_s$, and $I(p)$ is the image intensity at point $p \in \mathcal{I}$. The parameter \textbf{$\omega > 0$} controls the smoothness of the path and is also used to avoid the term of (\ref{potential}) in parentheses being $0$. The term in parentheses emphasizes voxels that differ from $p_s$. Voxels with image intensity similar to the intensity in $p_s$ neighborhood are assigned low values, whereas voxels from other regions, including boundaries, receive high values. The second term of (\ref{potential}) is a function of $d$, which is the distance map computed on the edge image. To get the edge image, we first compute the edge detector using the function
\begin{equation}
    g(|\nabla I|) = \frac{1}{1+K|\nabla I|^2} \label{Per_Mal}
\end{equation}
where $K > 0$. The function (\ref{Per_Mal}) is introduced in \cite{Perona_Malik_1990} for edge detection using anisotropic diffusion. A small value of the function (\ref{Per_Mal}) means weak diffusion, so the edges are preserved. The parameter $K$ controls edge sensitivity. Therefore, the edge image is a binary image $B$ obtained by thresholding ($\delta > 0$)

\begin{numcases}{}
g\!\bigl(\lvert \nabla I(p)\rvert\bigr) < \delta \;\implies\; B(p)=1, \label{eq:cond1}\\
g\!\bigl(\lvert \nabla I(p)\rvert\bigr) \ge \delta \;\implies\; B(p)=0. \label{eq:cond2}
\end{numcases}
The voxels $~p \in \mathcal{I}~$ for which (\ref{eq:cond1}) holds are called edge voxels, and the ones for (\ref{eq:cond2}) are called background voxels. The distance map sets the value of edge voxels to $0$ and assigns the minimal distance to the closest edge point to the background voxels. Therefore, points located at the aorta centerline get a value approximately equal to the aorta radius. When the edges are perfectly detected, we obtain a centered path within the aorta in one step. However, in non-enhanced CT data, several regions of the aorta may exhibit weak or missing edges, and the edge image may not capture them. Therefore, combining these two terms yields globally small values within the aorta and large values at the edges. We used a similar potential function in a previous work \cite{Konan_2024}
\begin{equation}
\Tilde{P}_{prev}(p) = \omega + \frac{|\Tilde{I}(p_s) - I(p)|}{\underset{q \in \mathcal{I}}{max}(|\Tilde{I}(p_s) - I(q)|)} \frac{1}{1 + d(p)}, ~\forall p \in \mathcal{I}, ~p \neq p_s  \label{potential_previous}
\end{equation}
Here, $\omega$ is added to the product of the two terms. The difference between the two potentials relies on the fact that, in this work, the smoothing effect of $\omega$ is reduced to favor the centering term (the term with distance). This change is essential because we have only two input points (in this work) and aim to extract a path within the aorta in a single step.

\bigskip

\subsection{Minimal path approach with keypoints detection}\label{minimal_path_approach_key_point_detection}

During the front propagation, when there are missing or weak edges, the front may flow outside the aorta, and other parts of the front may reach the specified final point before the front located within the aorta (see Fig.\ref{results_front_evolution}). This is an illustration of a shortcut, which occurs more often when the input data has low contrast, such as in non-enhanced CT images. One solution to overcome such shortcuts is to use the minimal path approach with keypoint detection \cite{Benmansour_2009}. The idea is to update the starting point during the front propagation, denoted as a keypoint along the path, before the front reaches the final point. In fast marching front propagation, the computation of the action map $U_{p_s}$ starts in the neighbourhood of the starting point $p_s$, and $U_{p_s}$ is computed progressively for all the domain points. A keypoint is the first point $p_k$ reached by the front for which 
\begin{equation}
    || p_s-p_k ||\geq \lambda 
    \label{algo_key_points}
\end{equation}
where $\lambda$ is a positive real number. The newly detected keypoint is defined as the new starting point. The process is repeated until the front reaches the final point (see Fig.\ref{front_evolution_keyp}). As the authors of \cite{Benmansour_2009} noted, the keypoint should lie within the aorta because the front propagates faster where the potential is small, i.e., along the aorta's centerline. Therefore, for a starting point inside the aorta and for an optimal $\lambda$, keypoints are found inside the aorta. Too small $\lambda$ leads to the detection of many keypoints. Too large $\lambda$ (larger than the distance between the aortic root and the descending aorta) may not solve the shortcut problem in the thoracic region. Therefore, the choice of $\lambda$ should consider the curved shape of the aortic arch, as it is the region where the shortcut occurs more frequently. Too large $\lambda$ means the front propagates longer before finding the new keypoint, which increases the chance to pass through a region with a missing edge, then a higher chance to find the keypoint outside. The optimal parameter $\lambda$ is found after an intensive search, as the aorta length and shape vary from one patient to another.  

\bigskip

\subsection{Numerical experiments}

In this section, we present path extraction using our proposed potential (\ref{potential}) in both artificial and real images. The objective is to highlight the improvements implied by $\Tilde{P}$ compared to potentials used in literature \cite{Deschamps_2001}, like
\begin{equation}
    P_{ref} = \omega + |\Tilde{I}(p_s) - I(p)|
    \label{reference_potential}
\end{equation}
where $\Tilde{I}(p_s)$ can be either the average image intensity value in a neighborhood of $p_s$, or the average value of image intensities in starting and final points, or the mean image intensity value in the region of interest (see \cite{Deschamps_2001}).

\subsubsection{Artificial data}

\begin{figure}[H]
\centering
\includegraphics[scale=0.24]{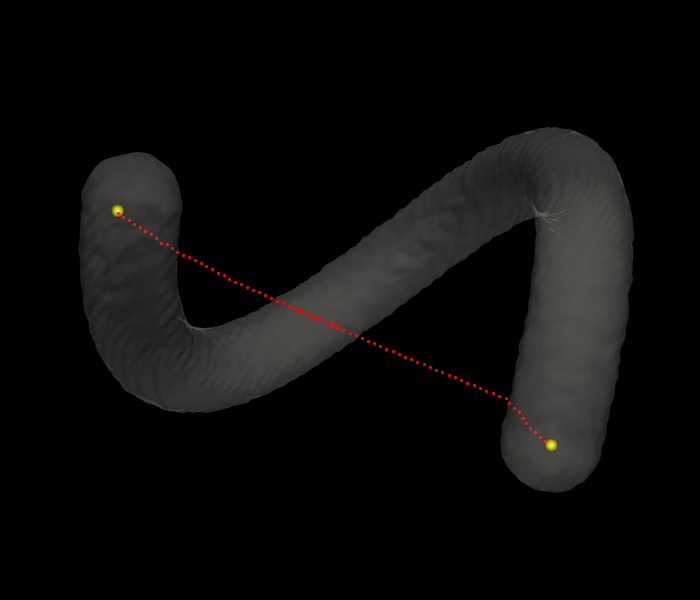}
\includegraphics[scale=0.24]{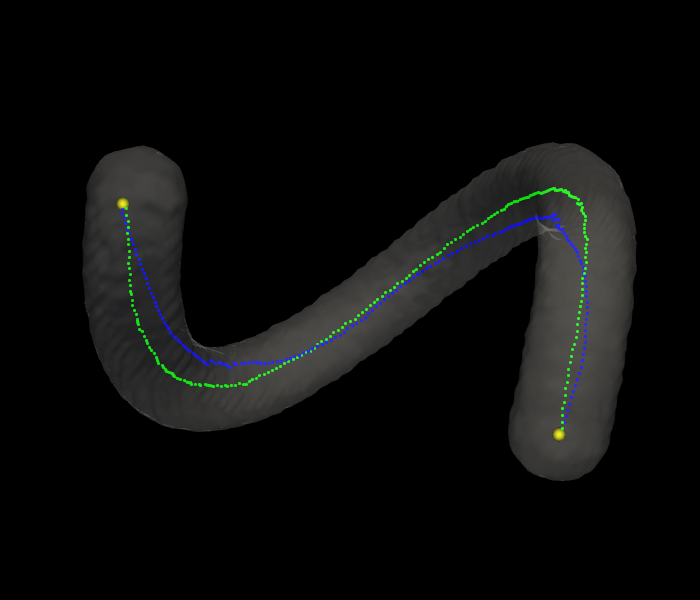}
\caption{Paths extraction within complex 3D shape. The yellow dots are the endpoints. The red path is a wrong path due to a shortcut. The blue path is located within the structure, but it is not centered, and the green path is centered.} 
\label{potential_artificial}
\end{figure}

In the original snake energy \cite{Kass_1988}, the location of the initial points is crucial to obtain the expected path. 
Energy minimization can become trapped in local minima, leading to a shortcut. The authors in \cite{Cohen_1997} introduced a new snake energy (\ref{snake_energy_minimal_path}) to achieve a global minimum, thereby reducing the likelihood of shortcuts. However, shortcuts may still occur when the potential $\Tilde{P}$ is not sufficiently contrasted or when the regularization parameter $\omega$ is too large (see the red curve in Fig.\ref{potential_artificial}). They suggested the tuning of $\omega$ to avoid shortcuts. In \cite{Deschamps_2001}, for tubular structures, shortcuts lead to a path that is tangent to the boundaries rather than centered (see blue curve in Fig.\ref{potential_artificial}). To obtain the centered path, they first segment the structure of interest and then use a distance-to-edge-based potential to derive the centered path. Our proposed potential $\Tilde{P}$ combines these two steps, resulting in a centered path when the edges are correctly extracted.  

\bigskip

In Fig.\ref{potential_artificial}, we present an example where the structure of interest is a 3D spiral-like structure. The object pixel values are set to $0.8$, and the background pixels are set to $0.7$. Using the potential $P_{ref}$, the regularization parameter $0 <\omega < 0.05$, and $\Tilde{I}(p_s)$ taken as the average voxel intensity value in the neighborhood of the starting point $p_s$, we obtain the blue path, which is not centered and tangent to the boundaries. When $\omega \geq 0.05$, the shortcut happens, and we obtain a wrong path (red). For $0 <\omega < 0.05$ and using $\Tilde{P}$ (\ref{potential}), we get a centered path (green) in a single step.

\begin{figure}[H]
\centering
\includegraphics[scale=0.24]{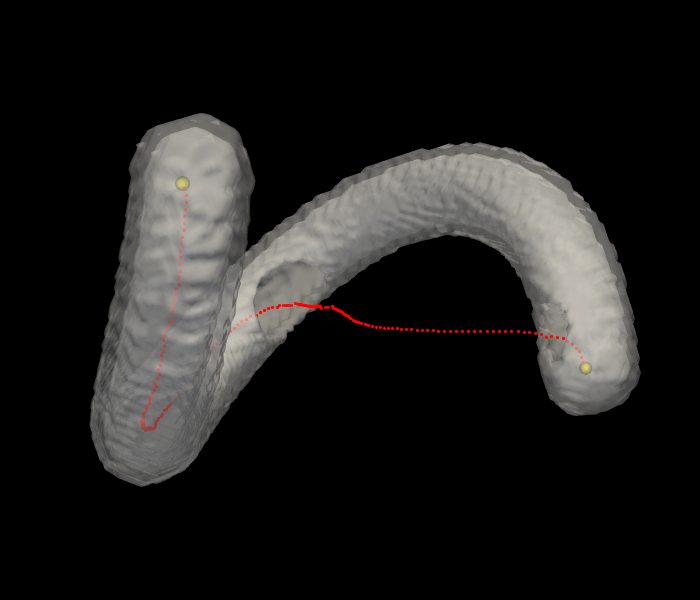}
\includegraphics[scale=0.24]{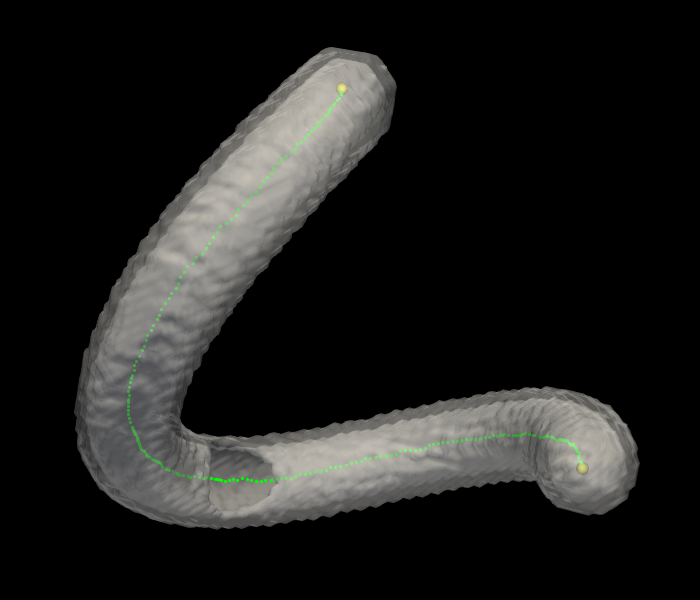}
\caption{Extracted path with missing edges near endpoints (red) and far away from the endpoints (green).} 
\label{illustrate_flowing_path}
\end{figure}

In Fig.\ref{illustrate_flowing_path}, we present the result of path extraction using the object in the previous example, but with parts of the object's boundaries missing. In the first case, we create a hole in the middle of the object and another hole close to the final point. The second has just one hole in the middle of the object. The distance to the edge term in $\Tilde{P}$ (\ref{potential}) decreases the potential value for all the voxels farther away from the object edges. Therefore, the front flows out of the tubular structure through the region with missing edges (hole), which can lead to a wrong path.  

The example of Fig.\ref{illustrate_flowing_path} demonstrates that we can still extract the expected path even when a few edges are missing and the endpoints are chosen within a region with complete edges. The case where edges are missing in many regions can be resolved using the growing approach. We will illustrate these cases using real data experiments. 

\subsubsection{Real data}

The datasets are provided by BIONT a.s. in Bratislava. The image data are acquired for patients suspected of having vasculitis. Each dataset comprises CT and PET 3D images in the same coordinate system. The CT image data are made of a sequence of 2D slices. The resolution of the slices is $512 \times 512$. In CT images, each voxel is assigned a numerical value (intensity) corresponding to the average X-ray attenuation within the voxel, expressed in Hounsfield Units (HU). Our datasets consist of data from six patients. Patients 1 and 2 correspond to data from the same patient, except that the scans were performed one year apart. In fact, the image origins and the number of slices in the 3D volumes differ, meaning that the images cannot be compared without primary processing. Considering the limited data available, we are treating them as two separate patients. The information about the CT data in our datasets is summarized in the following table

\begin{table}[H]
\centering
\caption{Information of CT data for patient $1$ (P1) to patient $6$ (P6).}
\renewcommand{\arraystretch}{1.5} 
\begin{tabular}{|c|c|c|c|c|c|c|c|}
\hline
\multicolumn{2}{|c|}{} & P 1 & P 2 & P 3  & P 4 & P 5 & P 6  \\ 
\hline
\multirow{2}{*}{Intensity (HU)} & min & $-1024$ & $-1024$ & $-3024$ & $-3024$ & $-1024$ & $-1024$ \\ 
\cline{2-8}
 & max & $2976$ & $2976$ & $3071$ & $3071$ & $3071$ & $3071$ \\ 
\hline
\multirow{3}{*}{Spacing (mm)} & $s_x$ & $1.172$ & $1.172$ & $0.977$ & $0.977$ & $0.977$ & $0.977$ \\ 
\cline{2-8}
 & $s_y$ & $1.172$ & $1.172$ & $0.977$ & $0.977$ & $0.977$ & $0.977$ \\ 
\cline{2-8}
 & $s_z$ & $2.5$ & $2.5$ & $2.5$ & $2.5$ & $1.5$ & $1.5$ \\ 
\hline
\end{tabular}
\label{spacing_patients_data}
\end{table}

To use similar parameters, we rescale the input data to the range $[0,1]$. We apply geodesic mean curvature flow filtering \cite{Kriva_2010} as a preprocessing step to remove noise while preserving edges. The endpoints for the path extraction are set manually. The threshold for the edge image is set to $0.2$ for patients 1 and 2, and to $0.25$ for patients 3 to 6. For strong edges, a small threshold is used to obtain the edge image. However, for weak edges, we need to increase the threshold to capture more edges. The regularization parameter $\omega$ is set proportional to the standard deviation of voxel intensities within a small neighborhood of the starting point $p_s$. 

\bigskip

The PET 3D data have lower resolution and are made of a sequence of 2D slices. The resolution of slices for Patients 1 and 2 is $144 \times 144$, for Patients 3 and 4 is $128 \times 128$, and for Patients 5 and 6 is $220 \times 220$. The information about the spacings is summarized in the following table

\begin{table}[H]
\centering
\caption{Information of PET data for patient $1$ (P1) to patient $6$ (P6).}
\renewcommand{\arraystretch}{1.5} 
\begin{tabular}{|c|c|c|c|c|c|c|c|}
\hline
\multicolumn{2}{|c|}{} & P 1 & P 2 & P 3  & P 4 & P 5 & P 6  \\ 
\hline
\multirow{3}{*}{Spacing (mm)} & $s_x$ & $4$ & $4$ & $3.91$ & $3.91$ & $3.3$ & $3.3$ \\ 
\cline{2-8}
 & $s_y$ & $4$ & $4$ & $3.91$ & $3.91$ & $3.3$ & $3.3$ \\ 
\cline{2-8}
 & $s_z$ & $4$ & $4$ & $4.25$ & $4.25$ & $3$ & $3$ \\ 
\hline
\end{tabular}
\label{spacing_patients_data_pet}
\end{table}

\begin{figure}[H]
\centering
\includegraphics[scale=0.15]{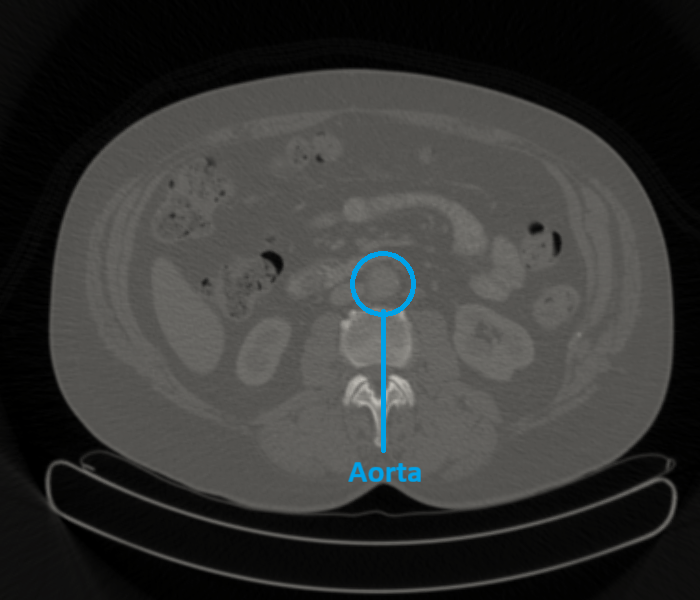}
\includegraphics[scale=0.15]{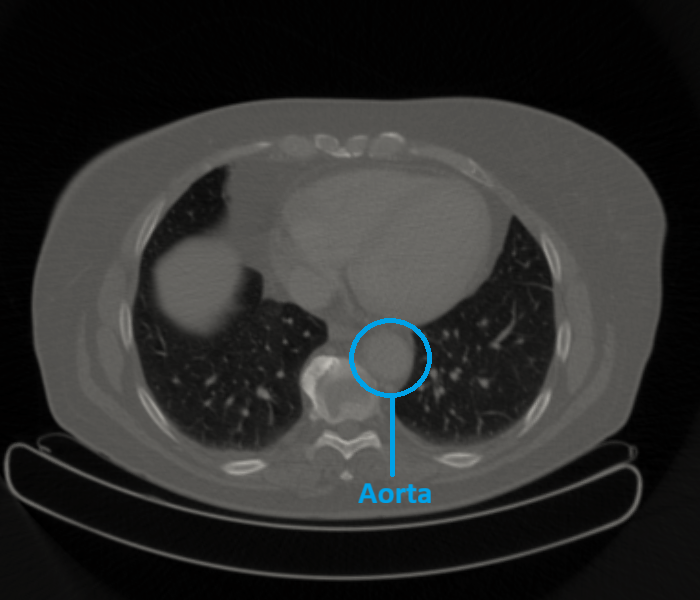}
\includegraphics[scale=0.15]{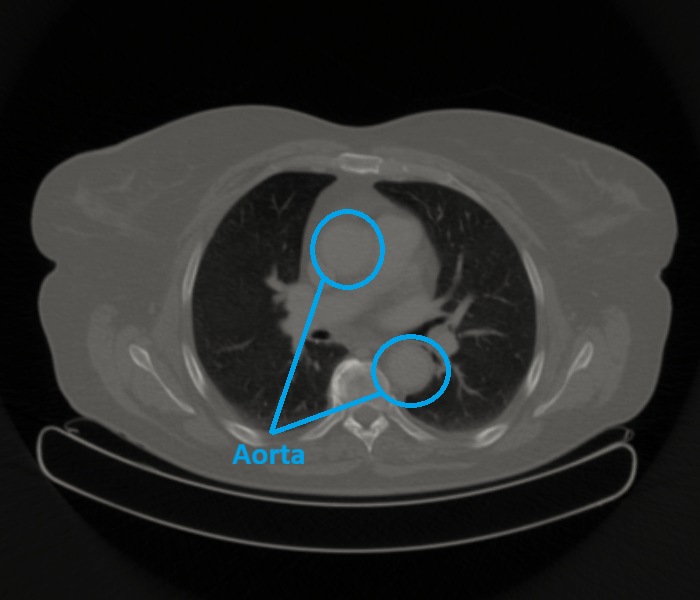}
\includegraphics[scale=0.15]{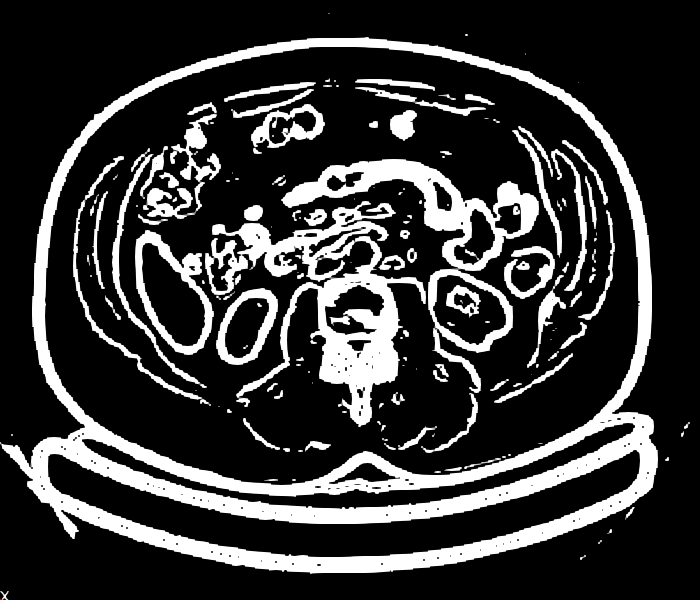}
\includegraphics[scale=0.15]{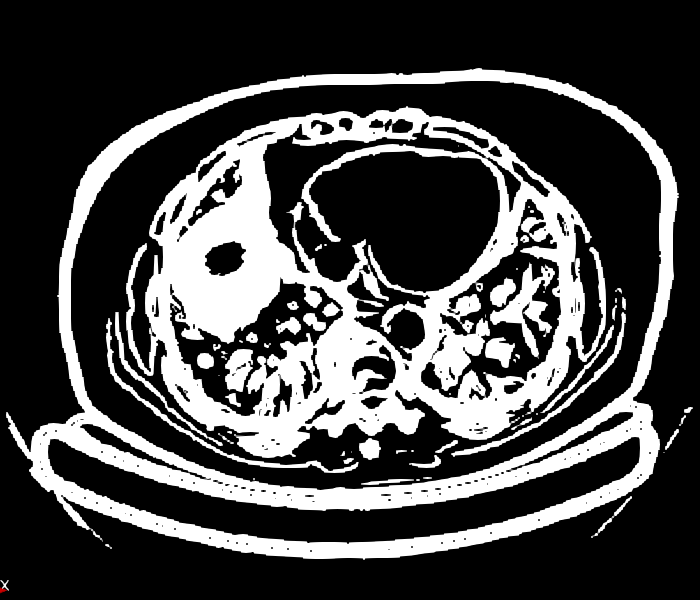}
\includegraphics[scale=0.15]{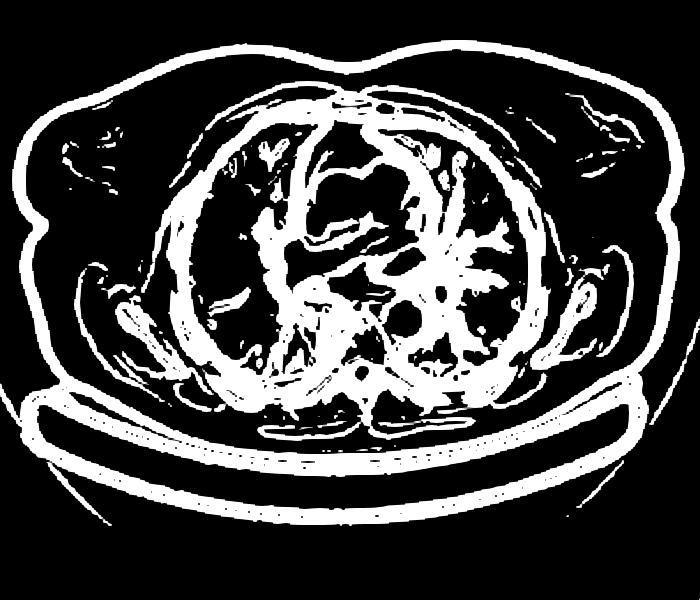}
\includegraphics[scale=0.15]{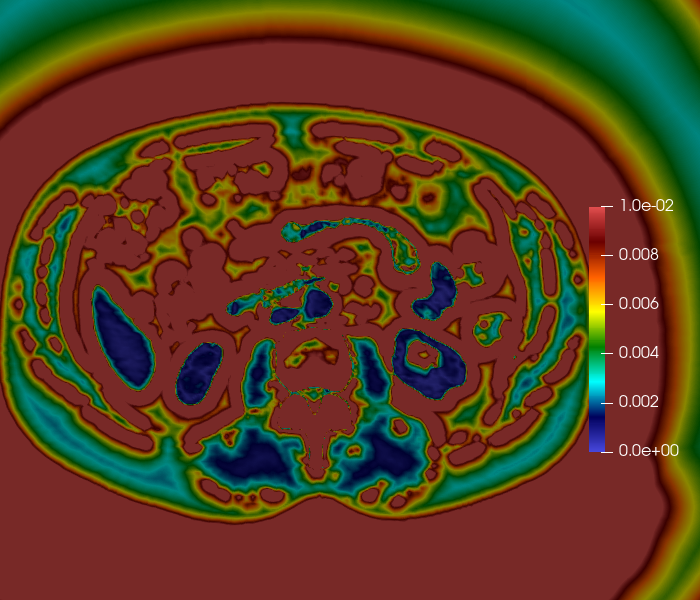}
\includegraphics[scale=0.15]{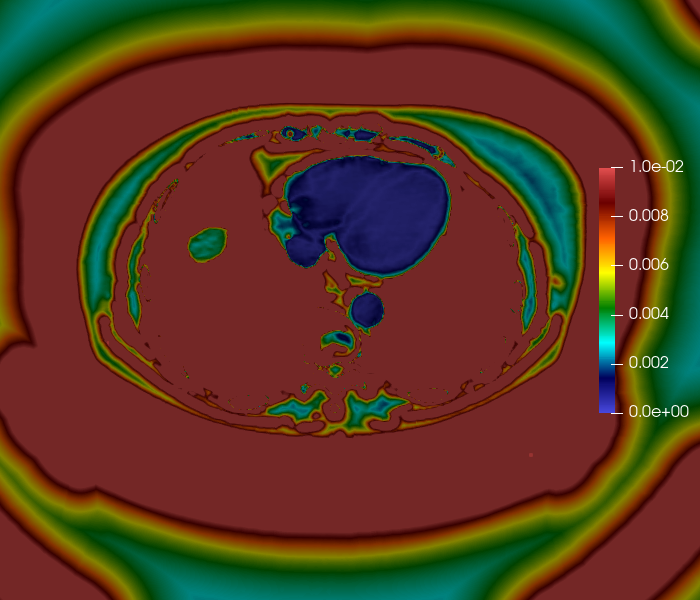}
\includegraphics[scale=0.15]{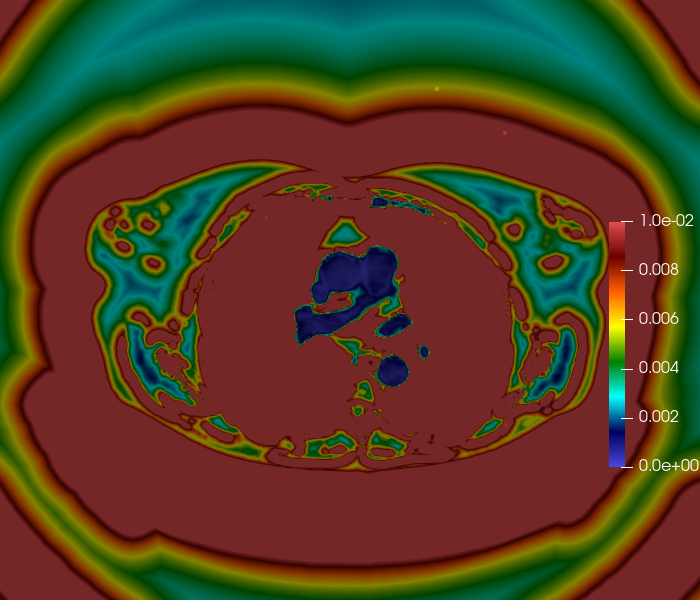}
\captionsetup{skip=10pt}
\caption{The first row shows three different axial views of the CT images where the aorta is visible (red circles). The second row shows the edge image computed from the previous input slices.  The third row shows the potential function corresponding to each slice. The slices are extracted from the 3D volume of Patient 1.} 
\label{illustrate_potential_real_image}
\vspace{4pt}
\end{figure}

\newpage
Fig.\ref{illustrate_potential_real_image} illustrates how the proposed potential $\Tilde{P}$ (\ref{potential}) values are small in the aorta region. This is our expectation to propagate the front inside the aorta before the other regions.

\begin{figure}[H]
\centering
\includegraphics[width=0.45\linewidth]{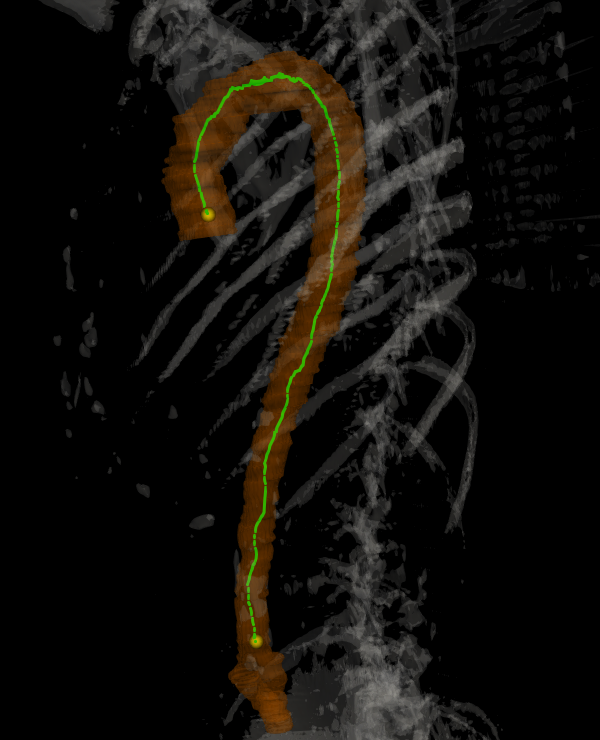}
\includegraphics[width=0.45\linewidth]{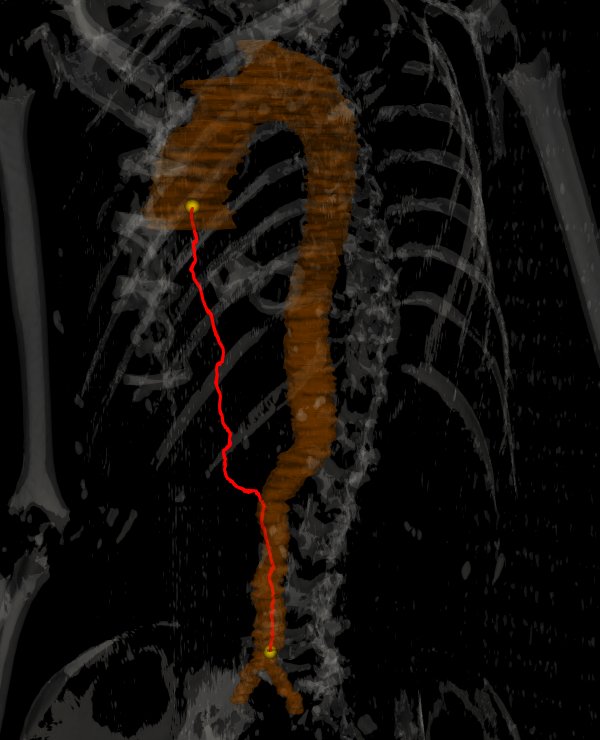}
\captionsetup{skip=10pt}
\caption{Extracted paths for patients 4 and 5. The aorta (orange) is segmented manually for visualisation purposes. The yellow dots are the starting and final points. The green path is the expected path for patient 4 using the potential $\Tilde{P}$, and the red path is an incorrect path for patient 5 due to a shortcut.} 
\label{results_path_extraction}
\end{figure}

\newpage
Fig.\ref{results_path_extraction} illustrates the extracted paths in a single step for two patients using the proposed potential $\Tilde{P}$ (\ref{potential}). For the potential $P_{ref}$ and various sets of parameters $\omega$, $\Tilde{I}(p_s)$, we were unable to obtain the expected path in one single step. However, when using the potential $\Tilde{P}$, the expected paths are extracted for four out of the six patients. That means the potential $\Tilde{P}$ improves the potential $P_{ref}$, which is defined with only voxel intensity values. The advantage of the proposed potential $\Tilde{P}$ is that it performs well even in regions with weak edges when the threshold for the edge image is adjusted. A thorough investigation into the patient data, which revealed incorrect paths, enables us to understand that these shortcuts occur in regions with high curvature and missing edges. To overcome this shortcut issue, we employ the growing minimal path approach to obtain the expected paths.

\begin{figure}[H]
\centering
\includegraphics[width=0.31\linewidth]{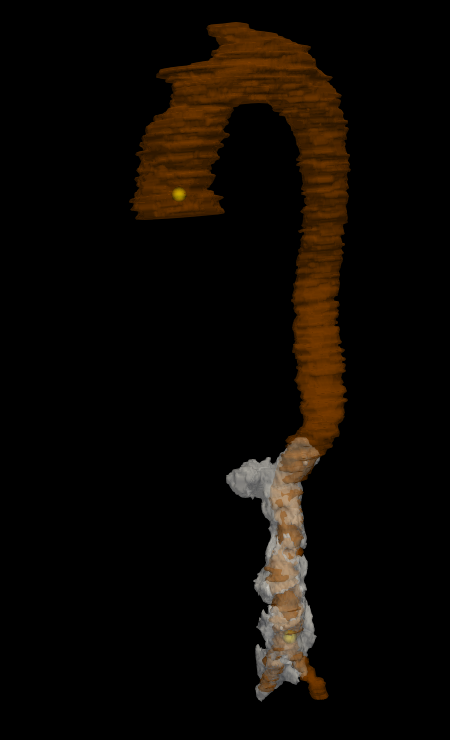}
\includegraphics[width=0.31\linewidth]{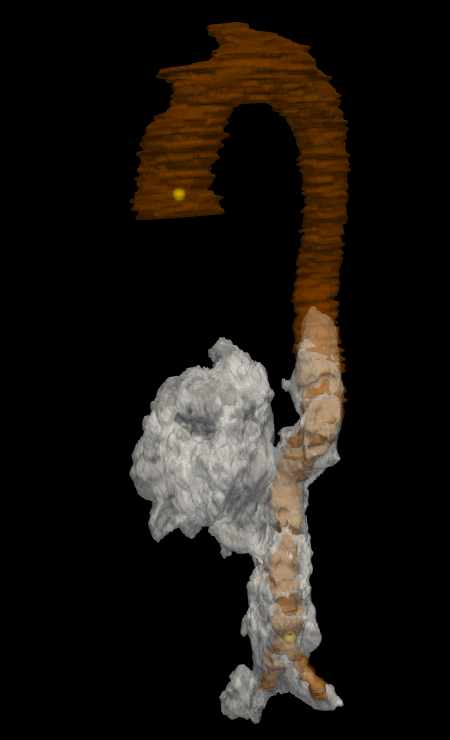}
\includegraphics[width=0.31\linewidth]{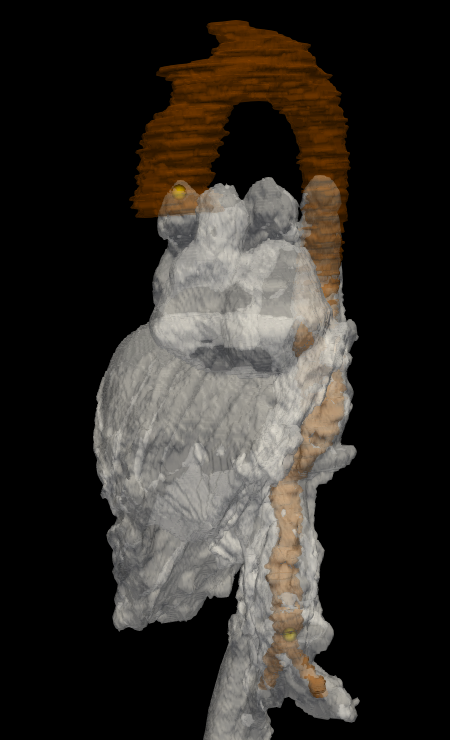}
\captionsetup{skip=10pt}
\caption{Three steps of the evolution of the front from the starting point to the final point for patient 5. It corresponds to the action map $U_{ps}$ used to extract the red path in Fig.\ref{results_path_extraction}.} 
\label{results_front_evolution}
\end{figure}

\newpage
The Fig.\ref{results_front_evolution} presents the evolution of the front from the starting point to the final point. Since the aorta is long with some missing edges, the front flows outside, reaching the final point before the front inside the aorta reaches the final point. This is a typical illustration of a shortcut when the potential is not sufficiently contrasted and the structure of interest is long.

\begin{figure}[H]
\centering
\includegraphics[width=0.26\linewidth]{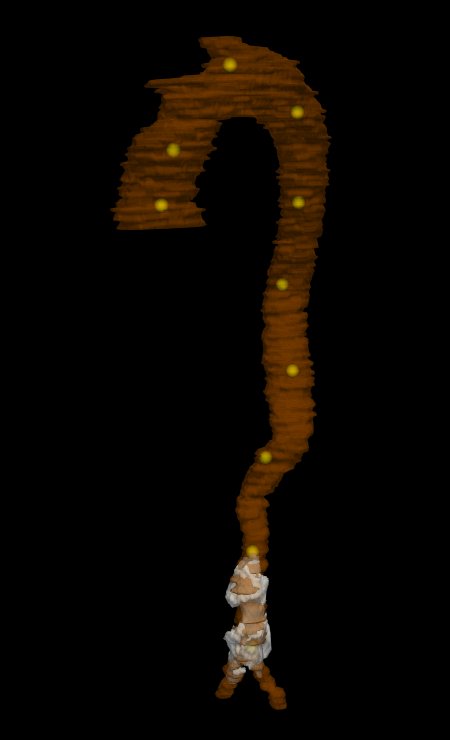}
\includegraphics[width=0.26\linewidth]{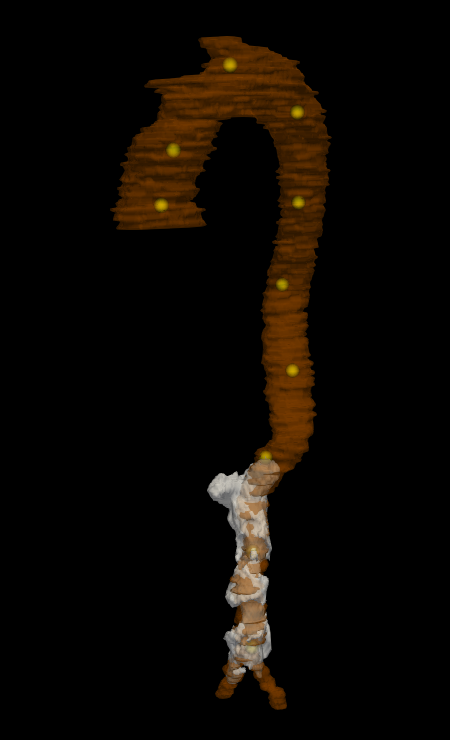}
\includegraphics[width=0.26\linewidth]{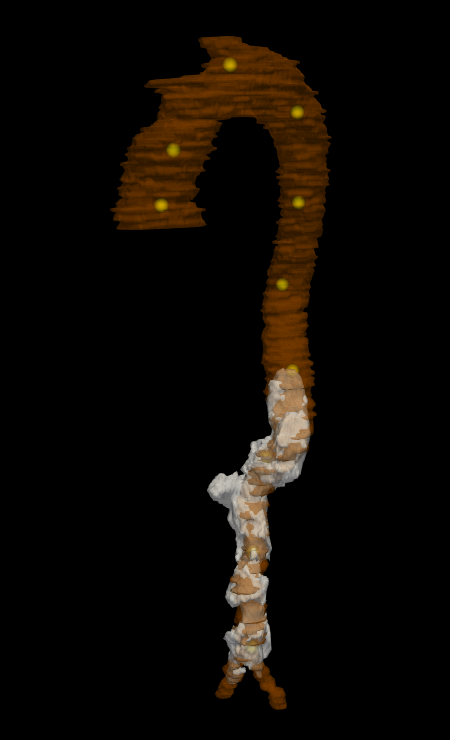}
\includegraphics[width=0.26\linewidth]{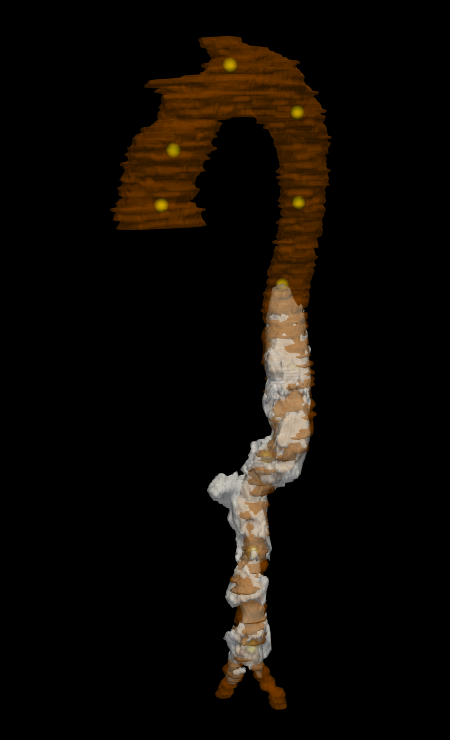}
\includegraphics[width=0.26\linewidth]{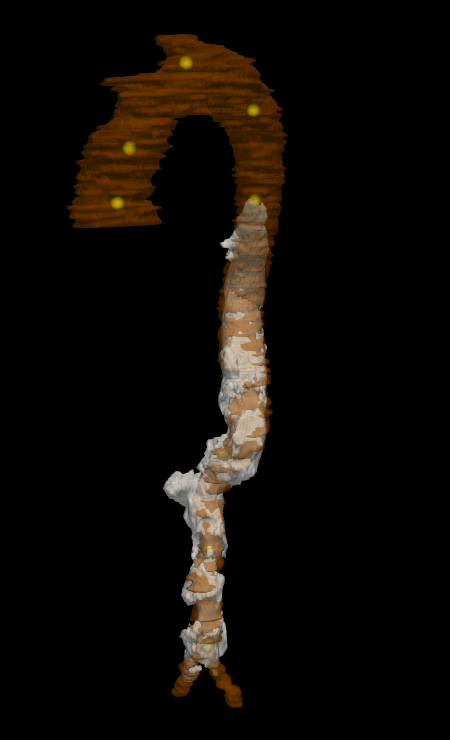}
\includegraphics[width=0.26\linewidth]{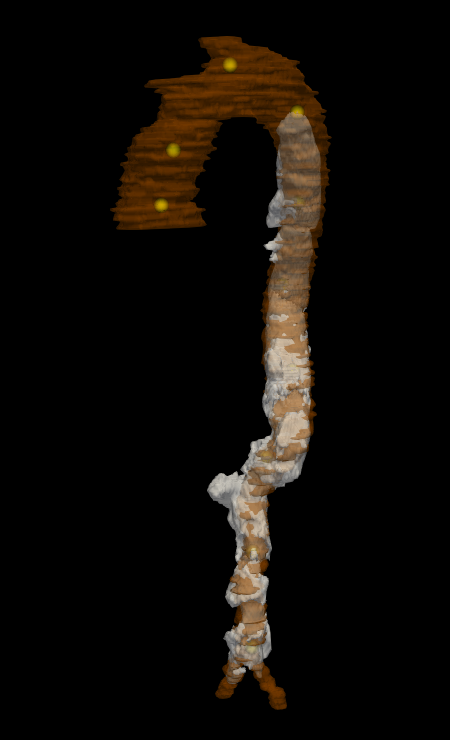}
\includegraphics[width=0.26\linewidth]{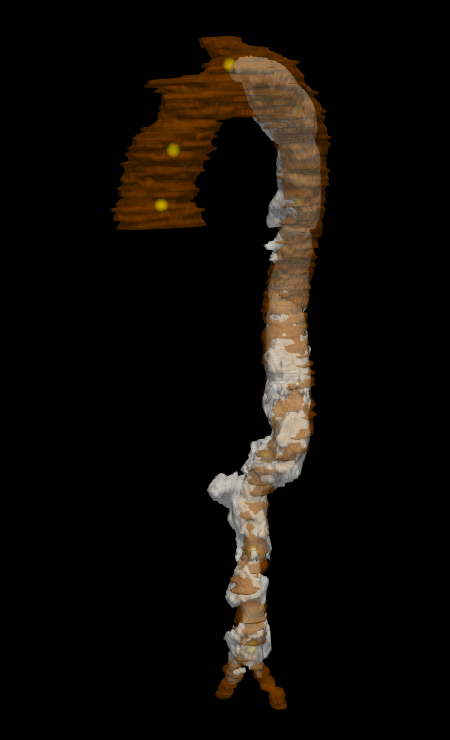}
\includegraphics[width=0.26\linewidth]{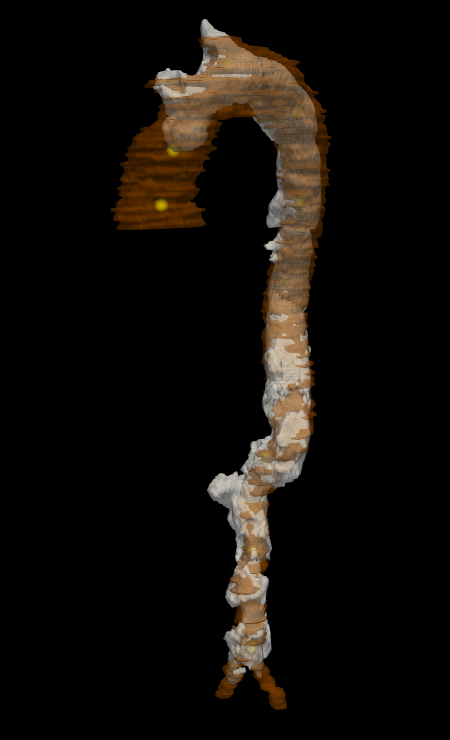}
\includegraphics[width=0.26\linewidth]{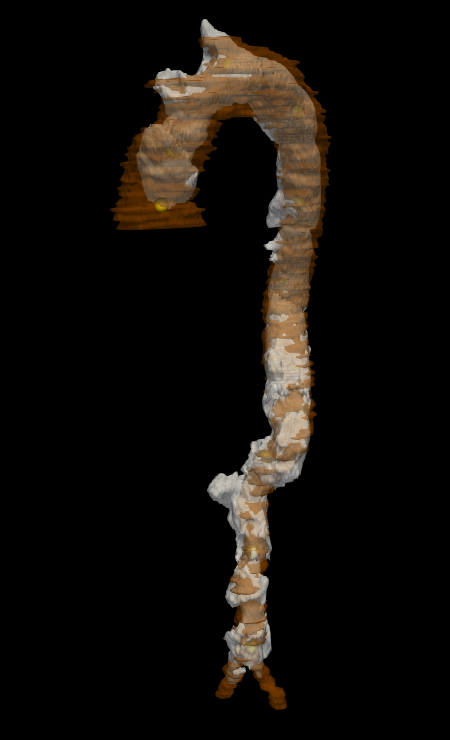}
\captionsetup{skip=10pt}
\caption{Front propagation with keypoints detections for patient 5. The detected keypoints are the yellow dots and $\lambda = 50$.} 
\label{front_evolution_keyp}
\end{figure}

\newpage
Fig.\ref{front_evolution_keyp} shows the front evolution with keypoint detection for $\lambda = 50$.

\begin{figure}[H]
\centering
\includegraphics[width=0.29\linewidth]{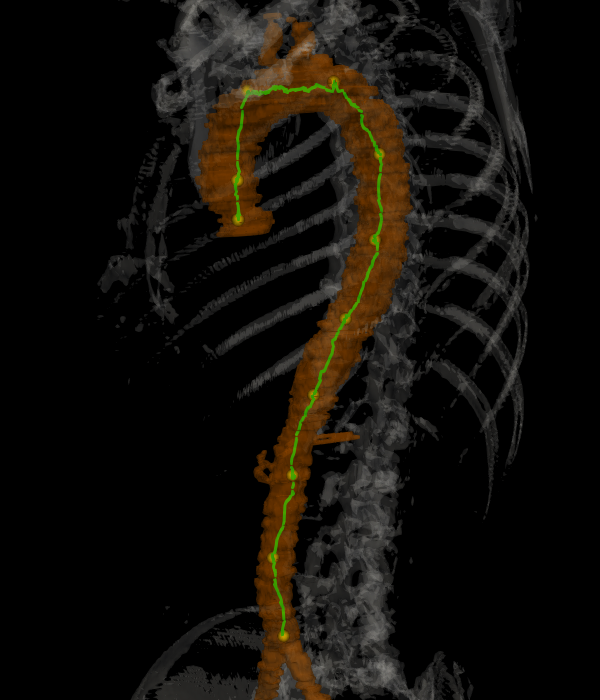}
\includegraphics[width=0.29\linewidth]{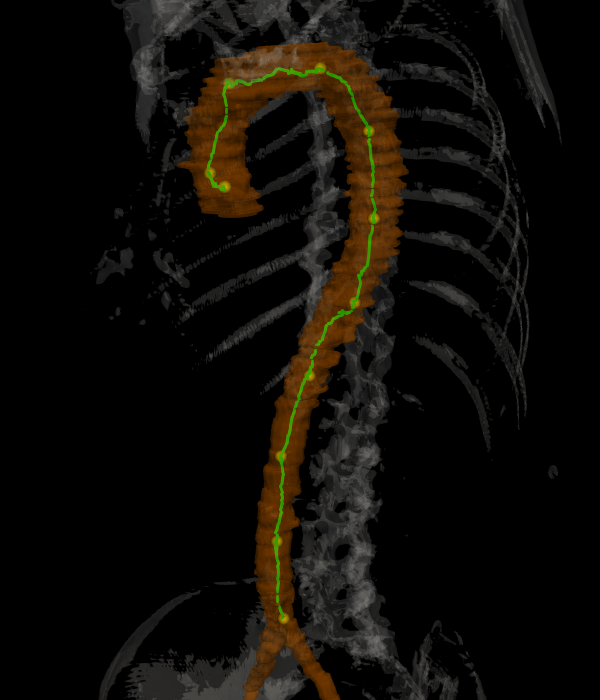}
\includegraphics[width=0.29\linewidth]{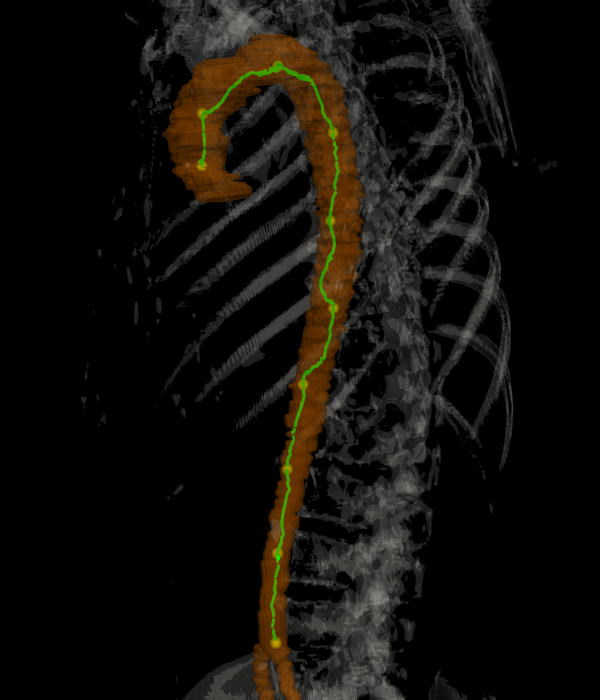}
\includegraphics[width=0.29\linewidth]{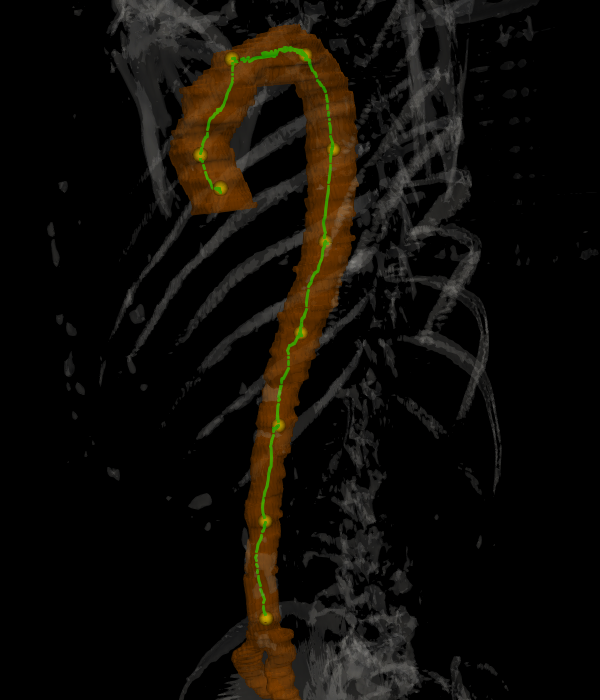}
\includegraphics[width=0.29\linewidth]{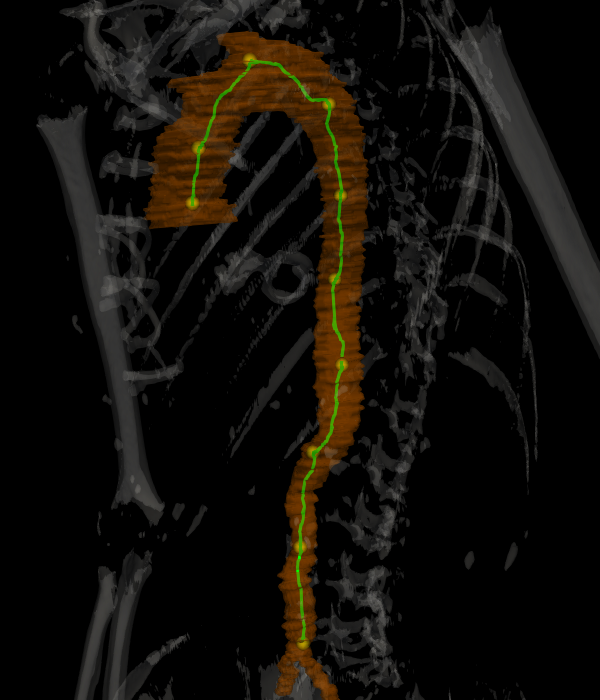}
\includegraphics[width=0.29\linewidth]{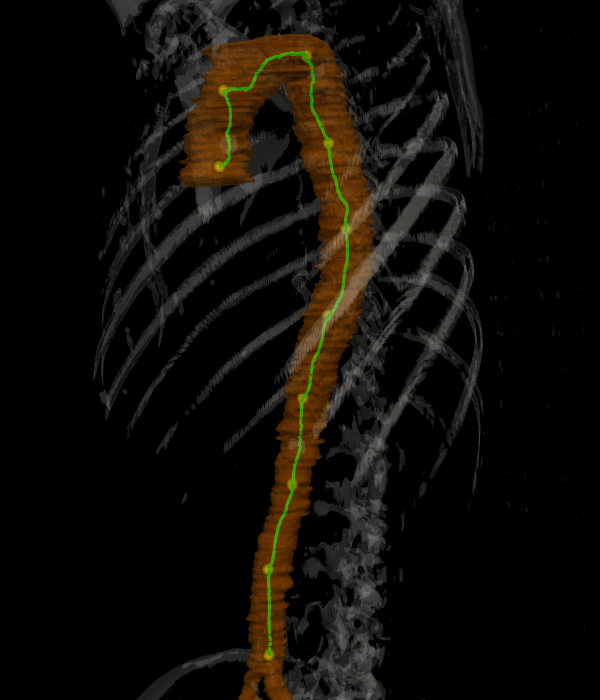}
\captionsetup{skip=10pt}
\caption{Extracted paths using the growing minimal path approach. Patients 1 to 3 in the first row and patients 4 to 6 in the second row. The yellow dots are the keypoints and $\lambda = 50$} 
\label{key_point_real_image}
\end{figure}

\newpage
The Fig.\ref{key_point_real_image} shows the extracted paths for our datasets using the minimal path approach with keypoint detection (see section \ref{minimal_path_approach_key_point_detection}). This approach allows us to overcome the shortcut issue observed earlier. All the keypoints are found within the aorta (for all patients) using the same $ \lambda = 50$. However, keypoints are sometimes found outside the aorta when the previous keypoint is close to an edge or in regions with weak or missing edges. For our dataset, this situation occurs more frequently in the heart region. Therefore, the challenge with this approach is how to define $\lambda$ to find all the keypoints within the aorta. The authors of \cite{DaChen_2014} introduced a solution to the problem. Their idea is to use a path score to select the candidate keypoint and use an anisotropic potential to keep finding keypoints in the target region. For the anisotropic potential, an anisotropic \textit{Fast Marching} method \cite{Mirebeau_2014} is required, which has computational complexity $\mathcal{O}(N~ln~N + N~ln~\kappa (\mathcal{M}))$ versus $\mathcal{O}(N~ln~N)$ for the isotropic (classical) \textit{Fast Marching} \cite{Sethian_1996}. The constant $N$ is the number of voxels in the image, $\kappa (\mathcal{M})$ the maximum anisotropy ratio, and $\mathcal{M}$ a symmetric definite positive matrix that encodes the desired anisotropy at a given point. Since our path extraction is a step in a comprehensive framework dedicated to real-time applications, we employ the original approach and search for the optimal $\lambda$ experimentally.

\section{Path centering}\label{path_centering}

The objective in the previous section was to extract a centered path in one single step to construct an appropriate initial condition for the GSUBSURF segmentation. Unfortunately, due to missing or weak edges, it is sometimes required to employ the growing minimal path approach to extract a path within the aorta. The extracted path is not centered due to the location of the keypoints, which are sometimes close to the edges (see the first image in row 2 of Fig.\ref{key_point_real_image}). In this section, we move the extracted path to the approximate centerline of the aorta. Our application is based on the 3D Lagrangean curve evolution presented by Mikula and Urbán in \cite{Mikula_Urban_2014}. The model moves a 3D curve smoothly in the normal direction within a complex tubular structure. 

\subsection{Mathematical model}

Let $~\Gamma: [0,1] \rightarrow \R^{3}$, $\Gamma = \{ \mathbf{r} (t, u), u\in[0,1]\}~$ be an open planar curve depending on time $t$ and let $~\mathbf{r}(t,u) = (x(t,u), y(t,u), z(t,u))~$ be a position vector of the curve $\Gamma$ for the parameter $u$ in time $t$. The discretized curves are represented by a sets of $3$D points $\textbf{r}_{i}^{n} = (x_{i}^{n}, y_{i}^{n}, z_{i}^{n})$, where $i = 0,...,m$, is the grid points number, and $n$ represents the time step. Our input curves are the paths extracted in the previous section (see Fig.\ref{key_point_real_image}). 

The motion of a curve in a general vector field $\mathbf{v}$ can be described by the following model
\begin{equation}
    \partial_{t} \mathbf{r} = \mathbf{v}(\mathbf{r}).
\end{equation}

For a given tubular structure, the distance to the edge inside the structure has a ridge along the centerline, and the gradient of that distance function points towards the ridge. In \cite{Mikula_Urban_2014}, the structure of interest is the colon, so the authors compute the distance map using the segmented colon. Contrary to what was done in \cite{Mikula_Urban_2014}, we do not have the segmented aorta as an input, so we use edge detection to extract the aorta edges. Then we threshold the edge detector to obtain an edge image where the aorta edges are visible (see Sect.\ref{define_potential}). Therefore, an edge image with "good enough" aorta edges is a good input for the distance function. Various approaches can be employed to compute the distance map, including \textit{Fast Sweeping} \cite{Zhao_2004}, \textit{time-relaxed Rouy-Tourin} \cite{Mikula_Urban_2014}, and \textit{Fast Marching} \cite{Sethian_1996}. \textit{Fast sweeping} is the quickest approach, but it is also the least precise. The \textit{time-relaxed Rouy-Tourin} and \textit{Fast Marching} show similar precision, but the \textit{Fast Marching} is much quicker. Therefore, we employ the \textit{Fast Marching} method, which solves
\begin{equation}
    |\nabla d| = 1
\end{equation}
to compute the distance map. The velocity vector field $\mathbf{v}$, for the curve evolution, is thus given by
\begin{equation}
    \mathbf{v} = \nabla d.
\end{equation}

The curve evolution model is based on the solution of the following partial differential equation \cite{Mikula_Urban_2014}
\begin{equation}
   \partial_t \mathbf{r} = \mu~\mathbf{N}_{\mathbf{v}} + \varepsilon~k\mathbf{N} + \alpha~\mathbf{T} 
   \label{curve_evolution}
\end{equation}
where $k\mathbf{N}$ is the curvature vector, $\alpha$ the tangential velocity , $\mathbf{T}$ the unit tangent vector to the curve, $\mu$ and $\varepsilon$ are parameters. The projection of the vector field $\mathbf{v}$ to the curve's normal plane is given by
\begin{equation}
    \mathbf{N}_{\mathbf{v}} = \mathbf{v} - (\mathbf{T} \cdot \mathbf{v})\mathbf{T}. 
    \label{normal_component}
\end{equation}
Since $\mathbf{T} = \partial_s \mathbf{r}$ and $k\mathbf{N} = \partial_{ss}\mathbf{r}$, the advection-diffusion form of (\ref{curve_evolution}) is given by
\begin{equation}
    \partial_t \mathbf{r} = \mu~\mathbf{N}_{\mathbf{v}} + \varepsilon~\partial_{ss}\mathbf{r} + \alpha~\partial_{s}\mathbf{r}. 
    \label{curve_evolution_ad_dif}
\end{equation}
The Dirichlet boundary condition accompanies the model (i.e, the first and last points are fixed). The normal component of the velocity vector field moves the curve points toward the ridge, and the tangential component redistributes uniformly the points along the curve \cite{Mikula_Urban_2014}.

\bigskip

Let us consider local orthogonal basis $(\mathbf{T}, \mathbf{N}_1, \mathbf{N}_2)$
\[
    \mathbf{N}_1 = \frac{\mathbf{N}_\mathbf{v}}{|\mathbf{N}_{\mathbf{v}}|},  \hspace{20pt}   \mathbf{N}_2 = \mathbf{N}_1 \times \mathbf{T}.
\]
If $|\mathbf{N}_{\mathbf{v}}| = 0$, then $\mathbf{N}_{1}$ is taken as the averaged value from the neighbouring grid points. The projections of the curvature vector $k\mathbf{N}$ onto $\mathbf{N}_1$ and $\mathbf{N}_2$ are given by
\[
    k_1 = k\mathbf{N} \cdot \mathbf{N}_1,  \hspace{20pt}   k_2 = k\mathbf{N} \cdot \mathbf{N}_2
\]
then the curvature vector satisfies $k\mathbf{N} = k_1\mathbf{N}_1 + k_2\mathbf{N}_2$ and the equation (\ref{curve_evolution_ad_dif}) can be written as \cite{Mikula_Urban_2014}
\begin{equation}
    \partial_t \mathbf{r} = U \mathbf{N}_1 + V \mathbf{N}_2 + \alpha \mathbf{T}
\end{equation}
where the normal components are given by
\[
    U = \varepsilon k_1 + \mu |\mathbf{N}_{\mathbf{v}}|,  \hspace{20pt}   V = \varepsilon k_2.
\]

The suitable tangential velocity $\alpha$ satisfies \cite{Mikula_Urban_2014}
\begin{equation}
    \partial_{s} \alpha = U k_1 + V k_2 - <U k_1 + V k_2 >_{\Gamma} + (\frac{L}{g} - 1) \omega_{r} 
    \label{tangential_velocity}
\end{equation}
where $g$ is the local segment length, $L$ the total length of the curve, $<\cdot>_{\Gamma}$ the arc-length average and $\omega_{r}$ is the redistribution rate. The choice of $\omega_{r}$ makes the quantity $\theta = ln(\frac{g}{L})$ converge to $0$, which means an asymptotically uniform spacing between curve points \cite{Mikula_Urban_2014}.

\subsubsection{Discretization}

The discretization follows the flowing finite volume method \cite{Mikula_Sevcovic_2001, Mikula_Sevcovic_2004, Mikula_Urban_2014}. Let $h_{i}^{n} = |\mathbf{r}_{i}^{n} - \mathbf{r}_{i - 1}^{n}|$ be the primal arc lengths and  $|\mathbf{r}^{n}_{i - \frac{1}{2}}- \mathbf{r}^{n}_{i + \frac{1}{2}}| = \frac{h^{n}_{i} + h^{n}_{i+1}}{2}$ the dual segment length at given time $n$. The time derivative $\partial_t \mathbf{r}$ is approximated by a forward finite difference and integrated over the dual segment (finite volume)
\begin{equation}
    \int_{\mathbf{r}^{n}_{i - \frac{1}{2}}}^{\mathbf{r}^{n}_{i + \frac{1}{2}}} \partial_t \mathbf{r}~ds =  \frac{h^{n}_{i+1} + h^{n}_{i}}{2}~\frac{\mathbf{r}_{i}^{n+1} - \mathbf{r}_{i}^{n}}{\tau} 
    \label{discret_time_derivative}.
\end{equation}
where $\tau$ is the time step.

The normal component of the velocity field $\mu\mathbf{N}_{\mathbf{v}}$ is approximated explicitly and integrated over the dual segment
\begin{equation}
     \int_{\mathbf{r}^{n}_{i - \frac{1}{2}}}^{\mathbf{r}^{n}_{i + \frac{1}{2}}} \mu~(\mathbf{N}_{\mathbf{v}})^{n}_{i} ~ds = \mu~\frac{h^{n}_{i+1} + h^{n}_{i}}{2} (\mathbf{N}_{\mathbf{v}})^{n}_{i} 
     \label{discret_normal}
\end{equation}

The curvature term $\varepsilon \partial_{ss}\mathbf{r}$ is approximated implicitly and integrated over the dual segment
\begin{equation}
    \int_{\mathbf{r}^{n}_{i - \frac{1}{2}}}^{\mathbf{r}^{n}_{i + \frac{1}{2}}} \varepsilon \partial_{ss}\mathbf{r}^{n+1}~ds = \varepsilon\left[\partial_{s}\mathbf{r}^{n+1} \right]_{\mathbf{r}^{n}_{i - \frac{1}{2}}}^{\mathbf{r}^{n}_{i + \frac{1}{2}}} = \varepsilon \left(\frac{\mathbf{r}^{n+1}_{i+1} - \mathbf{r}^{n+1}_{i}}{h^{n}_{i+1}} - \frac{\mathbf{r}^{n+1}_{i} - \mathbf{r}^{n+1}_{i-1}}{h^{n}_{i}}\right) 
    \label{discret_curvature}
\end{equation}

\bigskip

Integrating the advection term $\alpha \partial_{s}\mathbf{r}$ over the dual segment, we obtain 
\begin{equation}
    \begin{aligned}
        \int_{\mathbf{r}_{i - \frac{1}{2}}}^{\mathbf{r}_{i + \frac{1}{2}}} \alpha \partial_{s}\mathbf{r}~ds &= \alpha_i \left( \mathbf{r}_{i + \frac{1}{2}} - \mathbf{r}_{i - \frac{1}{2}} \right) = \frac{\alpha_i}{2}(\mathbf{r}_{i+1} - \mathbf{r}_{i-1}) \\
        &= - \frac{\alpha_i}{2}(\mathbf{r}_{i} - \mathbf{r}_{i + 1}) +  \frac{\alpha_i}{2}(\mathbf{r}_{i} - \mathbf{r}_{i - 1}) \label{advection_term}
    \end{aligned} 
\end{equation}

We consider the inflow velocity of advection $\alpha$ through the dual segment boundary $\mathbf{r}_{i - \frac{1}{2}}$ and outflow through $\mathbf{r}_{i + \frac{1}{2}}$ \cite{Mikula_2014_IIOE, Ambroz_2019}
\begin{equation}
    b^{in}_{i - \frac{1}{2}} = max(-\alpha_{i}, 0), \hspace{20pt} b^{out}_{i - \frac{1}{2}} = min(-\alpha_{i}, 0)
\end{equation}
\begin{equation}
    b^{in}_{i + \frac{1}{2}} = max(\alpha_{i}, 0), \hspace{20pt} b^{out}_{i + \frac{1}{2}} = min(\alpha_{i}, 0)
\end{equation}
then the advection term (\ref{advection_term}) can be written as
\begin{equation}
    -\frac{1}{2}(b^{in}_{i + \frac{1}{2}} + b^{out}_{i + \frac{1}{2}})(\mathbf{r}_{i} - \mathbf{r}_{i + 1}) - \frac{1}{2}(b^{in}_{i - \frac{1}{2}} + b^{out}_{i - \frac{1}{2}})(\mathbf{r}_{i} - \mathbf{r}_{i - 1})
    \label{adv_inflow_outflow}
\end{equation}

For the time discretization of the advection term, the inflow part is taken implicitly, and the outflow part explicitly, then (\ref{adv_inflow_outflow}) becomes
\begin{equation}
    - \frac{1}{2} b^{in}_{i + \frac{1}{2}} (\mathbf{r}_{i}^{n+1} - \mathbf{r}_{i + 1}^{n+1}) - \frac{1}{2} b^{in}_{i - \frac{1}{2}} (\mathbf{r}_{i}^{n+1} - \mathbf{r}_{i - 1}^{n+1}) - \frac{1}{2} b^{out}_{i + \frac{1}{2}} (\mathbf{r}_{i}^{n} - \mathbf{r}_{i + 1}^{n}) - \frac{1}{2} b^{out}_{i - \frac{1}{2}} (\mathbf{r}_{i}^{n} - \mathbf{r}_{i - 1}^{n}) 
    \label{advection_approx}
\end{equation}

\bigskip

The discrete curvature vector is given by
\begin{equation}
    (k\textbf{N})_{i}^{n} = \frac{2}{h_{i+1}^{n} + h_{i}^{n}}~(\frac{\textbf{r}_{i+1}^{n} - \textbf{r}_{i}^{n}}{h_{i+1}^{n}} - \frac{\textbf{r}_{i}^{n} - \textbf{r}_{i-1}^{n}}{h_{i}^{n}})
\end{equation}
and using the following approximation for the tangential unit vector
\begin{equation}
    \textbf{T}_{i}^{n} = \frac{\textbf{r}_{i+1}^{n} - \textbf{r}_{i}^{n}}{h_{i+1}^{n} + h_{i}^{n}}
\end{equation}
we compute the quantities $(\textbf{N}_1)_{i}^{n}$, $(\textbf{N}_2)_{i}^{n}~$, $k_{1i}^{n}~$, $k_{2i}^{n}~$, $U_{i}^{n}~$, and $V_{i}^{n}$. 

\bigskip

The approximation of the tangential velocity $\alpha^{n}_{i}$ is obtained by intergrating (\ref{tangential_velocity}) over the dual segment, and we obtain
\begin{equation}
    \alpha^{n}_{i} = \alpha^{n}_{i - 1} + h_{i}^{n}(U^{n}_{i} k^{n}_{1i} + V^{n}_{i} k^{n}_{2i}) - h_{i}^{n}<U k_1 + V k_2 >^{n}_{\Gamma} + (\frac{L^{n}}{m} - h_{i}^{n}) \omega_{r} 
\end{equation}
for $i = 1,\cdots, m-1$, setting $\alpha^{n}_{0} = 0$ and getting $\alpha^{n}_{m} = 0$.

\bigskip

The approximation of the PDE (\ref{curve_evolution_ad_dif}) by a flowing semi-implicit finite volume scheme is then obtained by combining the formulas \ref{discret_time_derivative}, \ref{discret_normal}, \ref{discret_curvature}, and \ref{advection_approx}. It can thus be written as
\begin{equation}
    \mathcal{A}_{i}^{n}~\mathbf{r}_{i-1}^{n+1} + \mathcal{B}_{i}^{n}~\mathbf{r}_{i}^{n+1} + \mathcal{C}_{i}^{n}~ \mathbf{r}_{i+1}^{n+1} = \mathcal{F}_{i}^{n} 
    \label{syst_path_centering}
\end{equation}
where the coefficients $\mathcal{A}_{i}^{n}$, $\mathcal{B}_{i}^{n}$, $\mathcal{C}_{i}^{n}$, and $\mathcal{F}_{i}^{n}$ are given by
\begin{equation}
\mathcal{A}_{i}^{n} = -\frac{\varepsilon}{h_{i}^{n}} - \frac{1}{2}b^{in}_{i - \frac{1}{2}} , \hspace{20pt} \mathcal{C}_{i}^{n} = -\frac{\varepsilon}{h_{i+1}^{n}} - \frac{1}{2}b^{in}_{i + \frac{1}{2}}
\end{equation}

\begin{equation}
    \mathcal{B}_{i}^{n} = \frac{h_{i}^{n} + h_{i+1}^{n}}{2 \tau} - (\mathcal{A}_{i}^{n} + \mathcal{C}_{i}^{n})
\end{equation}
\begin{equation}
\mathcal{F}_{i}^{n} = \frac{h_{i}^{n} + h_{i+1}^{n}}{2 \tau} \mathbf{r}_{i}^{n} + \mu (\mathbf{N}_{\mathbf{v}})_{i}^{n} \frac{h_{i}^{n} + h_{i+1}^{n}}{2} - \frac{1}{2} b^{out}_{i + \frac{1}{2}} (\mathbf{r}_{i}^{n} - \mathbf{r}_{i + 1}^{n}) - \frac{1}{2} b^{out}_{i - \frac{1}{2}} (\mathbf{r}_{i}^{n} - \mathbf{r}_{i - 1}^{n}) 
\label{discret_advection}
\end{equation}

The equation (\ref{syst_path_centering}) represents three tridiagonal systems for the $ x$, $ y$, and $ z$ coordinates of the grid points $\textbf{r}_{i}^{n}$, which represent the evolving curve points and can be solved by the Thomas algorithm. 

\newpage
\subsubsection{Results}
In this section, we present the results of the evolution of the extracted paths (all six patients) using the edge image to construct the velocity vector field $\mathbf{v}$. 

\begin{figure}[H]
\centering
\includegraphics[width=0.3\linewidth]{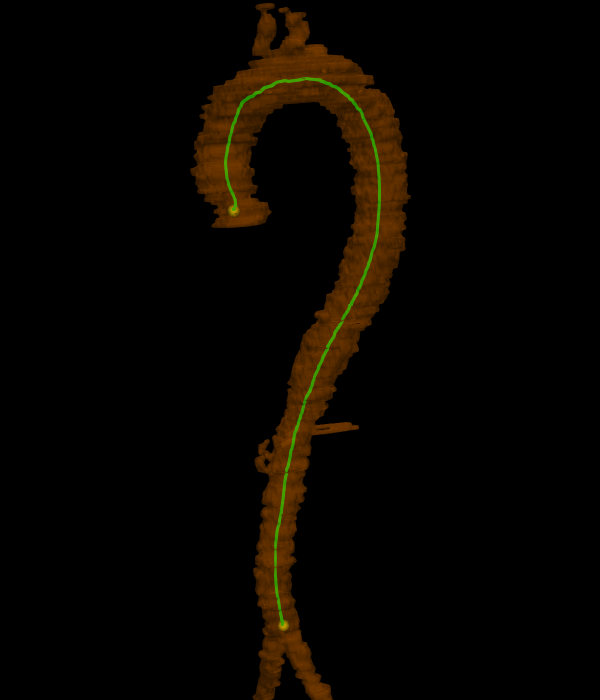}
\includegraphics[width=0.3\linewidth]{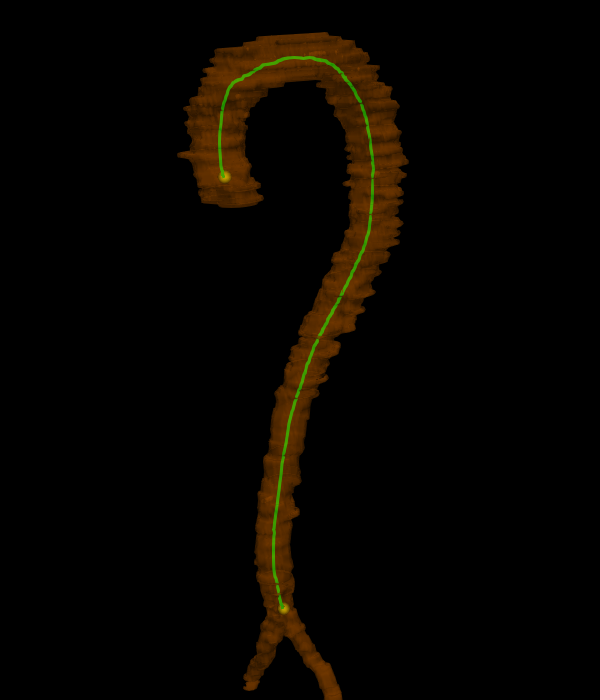}
\includegraphics[width=0.3\linewidth]{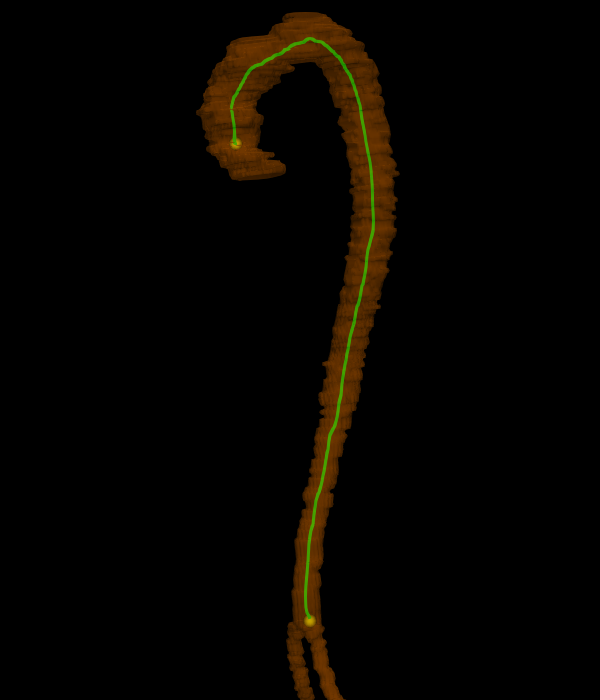}
\includegraphics[width=0.3\linewidth]{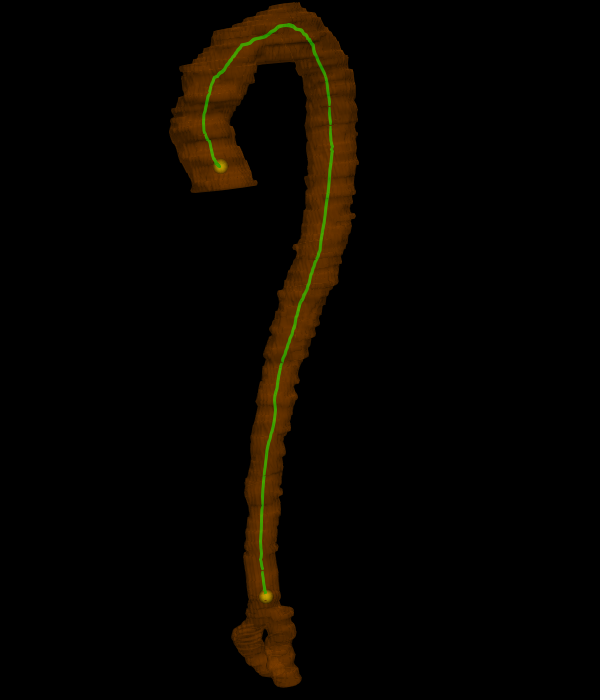}
\includegraphics[width=0.3\linewidth]{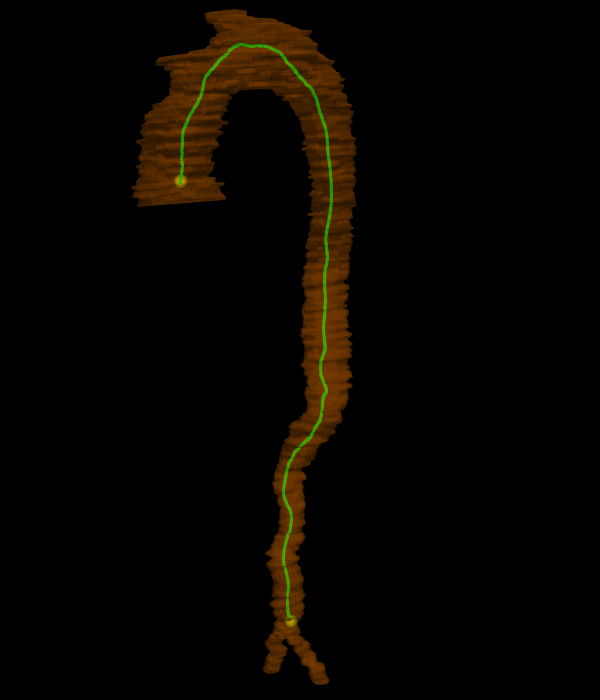}
\includegraphics[width=0.3\linewidth]{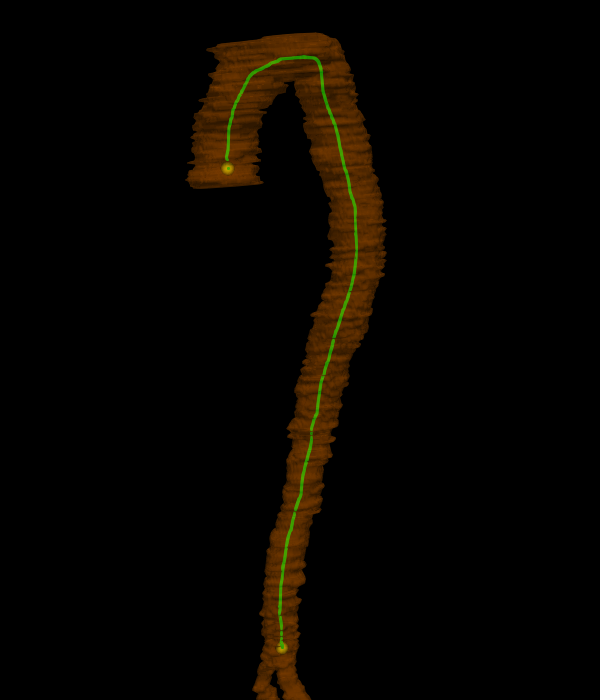}
\captionsetup{skip=10pt}
\caption{Centered paths (green) for six different patients. Patients 1 to 3 in the first row and patients 4 to 6 in the second row. The aorta (orange) is segmented manually for visualization purposes.} 
\label{results_pathcentering}
\end{figure}

Fig.\ref{results_pathcentering} shows the results of the curve evolution model applied to the paths extracted in the previous sections. We stop the evolution when the total curve length is no longer changing. We used the following parameters $\varepsilon = \mu = \omega_r = 1$. For the edge image, $K = 1000$ for all patients. The threshold is set to $\delta = 0.20$ for patients 1 and 2, and $\delta = 0.25$ for patients 3 to 6. When adjusting the threshold value for the edge image, it can be observed that the path is not perfectly centered in certain regions, where the edge information is not accurately captured. In contrast, the path remains well-centered in regions where the edge detection is reliable. These results demonstrate that the model \cite{Mikula_Urban_2014} exhibits robustness in handling incomplete or missing edge information within the structure of interest.  

\newpage
\section{The GSUBSURF model to segment the aorta}

Let $I : \mathcal{I} \subset \R^{3} \to \R$, represents the image to be segmented. The generalized subjective surface (GSUBSURF) model consists of defining an initial condition and letting it evolve according to the following advection-diffusion partial differential equation \cite{Mikula_Peyrieras_2007, Mikula_Peyrieras_2011, Mikula_2009}
\begin{equation}
u_{t} - w_{ad}~\nabla g \cdot \nabla u - w_{dif}~|\nabla u|~g ~\nabla \cdot \left( \frac{\nabla u}{|\nabla u|} \right) = 0 \label{gsubsurf}   
\end{equation}
where $u(.,t)$ is the segmentation function, $g(s) = \frac{1}{1 + Ks^{2}}$ is an edge detector function, $K > 0$ is used to control the edge detection sensibility and $s=|\nabla I|$. The image $I$ is filtered by the geodesic mean curvature flow filter to remove the noise and keep the edges \cite{Kriva_2010}. The parameters $w_{ad}, w_{dif} \in \R$ are used to control the advection and diffusion parts of the model. The velocity vector field $~\textbf{v} = -w_{ad}\nabla g~$ is important since it points toward the edges. The advection term moves the initial curve toward the edges, while the diffusion term handles the changes in the curve shape during evolution. We solve the problem in the domain $\mathcal{I} \times [0, T]$ and use the following boundary and initial conditions
\[
    u(p, t) = 0, \hspace{0.1cm} \forall p \in \partial \mathcal{I}, \hspace{0.5cm} \text{and} \hspace{0.5cm} u(p, 0) = u^{0}(p).
\]

\vspace{5pt}
\subsection{Initial segmentation function}

The initial segmentation function is typically defined by a peak located approximately at the center of gravity of the object to be segmented \cite{Mikula_2009, Mikula_Peyrieras_2007}. Therefore, we use the aorta's approximate centerline extracted in the previous section to construct the initial condition
\begin{equation}
    u^{0}(p) = \frac{1}{1 + d(p)} \label{initial_solution}, \hspace{0.2cm} p \in \mathcal{I} \subset \R^{3}
\end{equation}
where $d(p)$ is minimal distance between $p$ and the input curve $\Gamma$.

\vspace{5pt}
\subsection{Finite volume space discretization}
We consider a rectangular finite volume that corresponds to a voxel in the input image (see Table \ref{spacing_patients_data}). We denote $p$ the current finite volume and $N(p) = \{e,w,n,s,t,b\}$ the set of its neighbours. The finite volume size is defined by the spacing in each direction $h_x$, $h_y$, and $h_z$. Let $m(p)$ denote the volume of the finite volume $p$, $m(p) = h_xh_yh_z$. We denote by $e_{pq}$ the common boudary for two adjacent finite volumes $p$ and $q$, with erea $m(e_{pq}) = h_x h_y$ for the plane $(x,y)$,  $m(e_{pq}) = h_x h_z$ for the plane $(x,z)$ and  $m(e_{pq}) = h_y h_z$ for the plane $(y,z)$. The edge connecting the centers of the current finite volume $p$ and a given neighbour $q$ is denoted $\sigma_{pq}$, and $m(\sigma_{pq})$ measures the edge with $m(\sigma_{pq}) \in \{h_x, h_y, h_z\}$.

\bigskip

We denote $\mathbf{v} = -w_{ad} \nabla g$ and consider the transformation
\begin{equation}
    \nabla \cdot ( \textbf{v}  u ) = u \nabla \cdot \textbf{v} + \textbf{v} \cdot \nabla u  ~\Longrightarrow~  \textbf{v} \cdot \nabla u = \nabla \cdot ( \textbf{v}  u ) - u \nabla \cdot \textbf{v}~\label{transfo}
\end{equation}

Integrating (\ref{gsubsurf}) over the finite volume $p$, we get
\begin{equation}
    \underbrace{\int_{p}u_t~dp}_{\mathrm{(I)}} + \underbrace{\int_{p}\nabla \cdot ( \textbf{v}  u )dp - \int_{p}u \nabla \cdot \textbf{v}dp}_{\mathrm{(II)}} -  \underbrace{\int_{p} w_{dif} |\nabla u| g \nabla \cdot \left( \frac{\nabla u}{|\nabla u|} \right)dp}_{\mathrm{(III)}} = 0
\end{equation}

Defining the solution in finite volume $p$ as constant $u_p$ and applying Green's theorem, $\mathrm{(II)}$ can be written as
\begin{equation}
   \mathrm{(II)} \approx \sum_{q \in N(p)} \int_{e_{pq}} u_{pq}\mathbf{v}\cdot n_{pq}~ds - u_p  \sum_{q \in N(p)} \int_{e_{pq}} \mathbf{v}\cdot n_{pq}~ds 
\end{equation}
where $n_{pq}$ denotes the outer normal to the finite volume $p$ boundaries. The solution on the interface $e_{pq}$ is denoted by $u_{pq}$. The integrated flux through a voxel side is given by \cite{Mikula_2012}
\begin{equation}
   v_{pq} = \int_{e_{pq}} \textbf{v}\cdot n_{pq}~ds = \int_{e_{pq}} - w_{ad} \nabla g \cdot n_{pq}~ds \approx - w_{ad} G_{pq}m(e_{pq}).
\end{equation}
Using the integral fluxes, we define the inflows and the outflows through voxel sides \cite{Mikula_2014_IIOE, Mikula_2012, Mikula_2009}, then we get 
\begin{equation}
   \mathrm{(II)} \approx \sum_{q \in N(p)} min(v_{pq}, 0) (u_{q} - u_{p}).
\end{equation}
We apply Green's theorem on $\mathrm{(III)}$ and we get
\begin{equation}
    \mathrm{(III)} \approx w_{dif}~|\nabla u_p|~g_p \sum_{q \in N(p)} m(e_{pq})~\frac{1}{|\nabla u_{pq}|} \frac{u_{q} - u_{p}}{m(\sigma_{pq})}.
\end{equation}

\subsection{Time discretization}

\noindent The term $\mathrm{(I)}$ is approximated by forward finite difference
\begin{equation}
    \mathrm{(I)} \approx \frac{u_{p}^{n+1} - u_{p}^{n}}{\tau} m(p).
\end{equation}
The advection term $\mathrm{(II)}$ is approximated implicitly in time
\begin{equation}
    \mathrm{(II)} \approx \sum_{q \in N(p)} min(v_{pq},0) (u_{q}^{n+1} - u_{p}^{n+1}).
\end{equation}
The diffusion term $\mathrm{(III)}$ is approximated by a semi-implicit scheme
\begin{equation}
    \mathrm{(III)} \approx  w_{dif}~|\nabla u_{p}^{n}|~g_p \sum_{q \in N(p)} \frac{m(e_{pq})}{m(\sigma_{pq})} ~\frac{u_{q}^{n+1} - u_{p}^{n+1}}{|\nabla u_{pq}^{n}|}.
\end{equation}

\bigskip
Combining the terms $\mathrm{(I)}$, $\mathrm{(II)}$ and $\mathrm{(III)}$, we get the full discretization of (\ref{gsubsurf}) 
\begin{multline}
    \frac{u_{p}^{n+1} - u_{p}^{n}}{\tau} m(p) + \sum_{q \in N(p)} min(v_{pq},0) (u_{q}^{n+1} - u_{p}^{n+1}) = \\ 
    w_{dif}~|\nabla u_{p}^{n}|~g_p \sum_{q \in N(p)} \frac{m(e_{pq})}{m(\sigma_{pq})} ~\frac{u_{q}^{n+1} - u_{p}^{n+1}}{|\nabla u_{pq}^{n}|}
\end{multline}
which is a linear system with unknowns $u_{p}^{n+1}$. We solve it by the successive over-relaxation (SOR) method.

The Evans-Spruck regularization \cite{Mikula_2009} is used to prevent possible zero gradients in the denominators
\begin{equation}
| \nabla u^{n}_{pq} | \approx | \nabla u^{n}_{pq} |_{\varepsilon} = \sqrt{\varepsilon^2 + |\nabla u^{n}_{pq}|^2}, \text{  } \varepsilon > 0
\end{equation}

We use the reduced diamond-cell approximation \cite{Mikula_2009} to compute the norm of the gradient $|\nabla u^{n}_{pq}|_{\varepsilon}$ and $|\nabla I|_q$. The terms $g_p$  and $|\nabla u^n_p|_{\varepsilon}$ are obtained using average value of $|\nabla I|_q$ and $|\nabla u^n_{pq}|$ in neighbouring finite volume $q$

\[
g_p = g(\frac{1}{6} \sum_{q \in N_p} |\nabla I|_q), \hspace{0.5cm} |\nabla u^{n}_{p}|_{\varepsilon} = \sqrt{\frac{1}{6}\sum_{q \in N_p } |\nabla u^{n}_{pq}|^{2} + \varepsilon^{2}}
\]

We define an approximation of the gradient of the edge detector $g$ using central differences \cite{Mikula_Urban_2014}: $\nabla g_{p} = (G_{pe}, G_{pn}, G_{pt}) = (-G_{pw}, -G_{ps}, -G_{pb})$, where 
\[
    G_{pe} = -G_{pw} \approx \frac{g_e - g_w}{2h_x}, \hspace{0.25cm} G_{pn} = -G_{ps} \approx \frac{g_n - g_s}{2h_y}, \hspace{0.25cm} G_{pt} = -G_{pb} \approx \frac{g_t - g_b}{2h_z}
\]

Another approach to approximating the gradient of the edge detector is suggested in \cite{Mikula_2009, Mikula_Peyrieras_2007}, which is particularly convenient for sharp edges. The current choice is convenient for images with weak or missing edges. 

\subsection{Results}

\begin{figure}[H]
\centering
\includegraphics[width=0.3\linewidth]{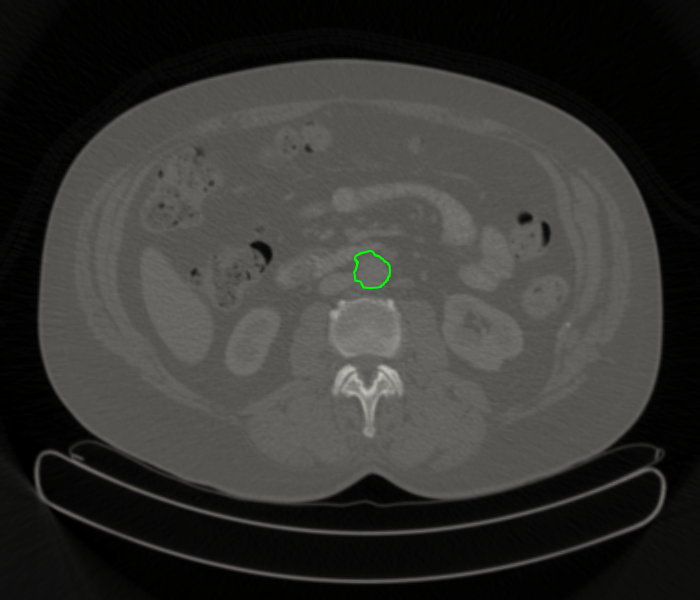}
\includegraphics[width=0.3\linewidth]{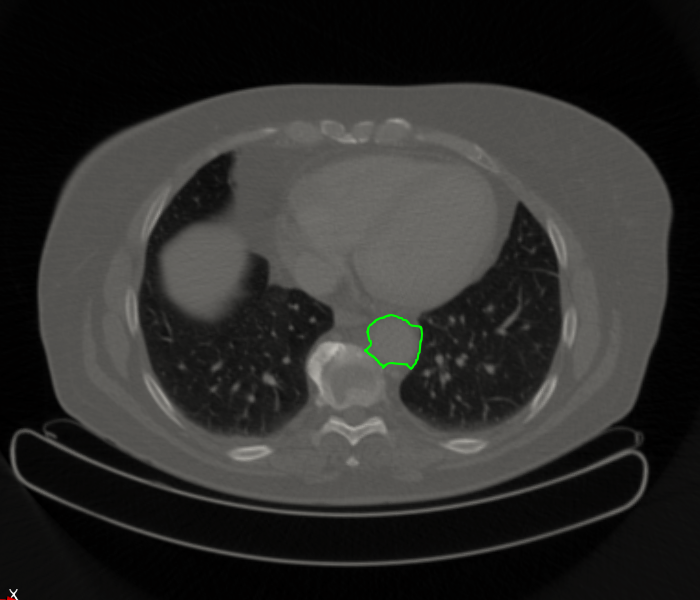}
\includegraphics[width=0.3\linewidth]{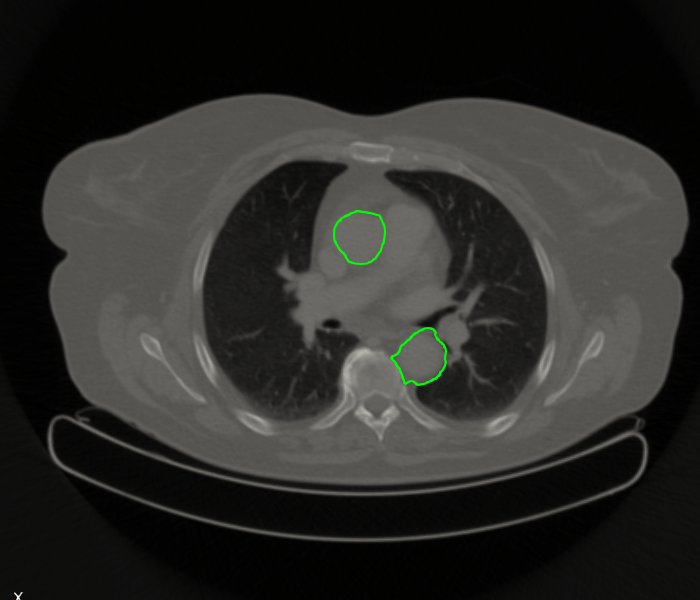}
\captionsetup{skip=10pt}
\caption{Segmentation result, 2D view (axial view) at three aorta regions in patient 1 image data.} 
\label{Result_g_subsurf_2D}
\end{figure}

In this section, we present the segmentation results of the aorta by the GSUBSURF model for the entire dataset. Fig.\ref{Result_g_subsurf_2D} presents the segmentation result for patient 1, in 2D views. The slices are chosen from the abdominal and the thoracic aorta.

\begin{figure}[H]
\centering

\begin{subfigure}[b]{0.3\textwidth}
    \centering
    \includegraphics[width=1.0\linewidth]{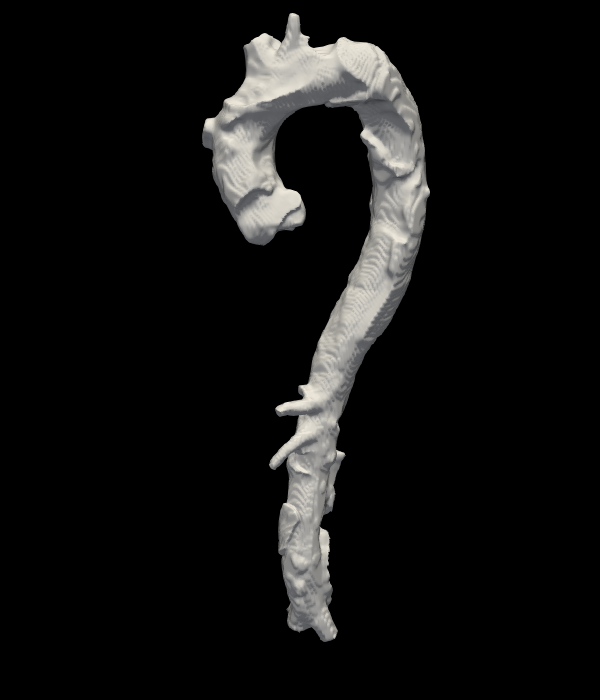}
    \caption{Patient 1, isoline $0.2$.}
\end{subfigure}
\begin{subfigure}[b]{0.3\textwidth}
    \centering
    \includegraphics[width=1.0\linewidth]{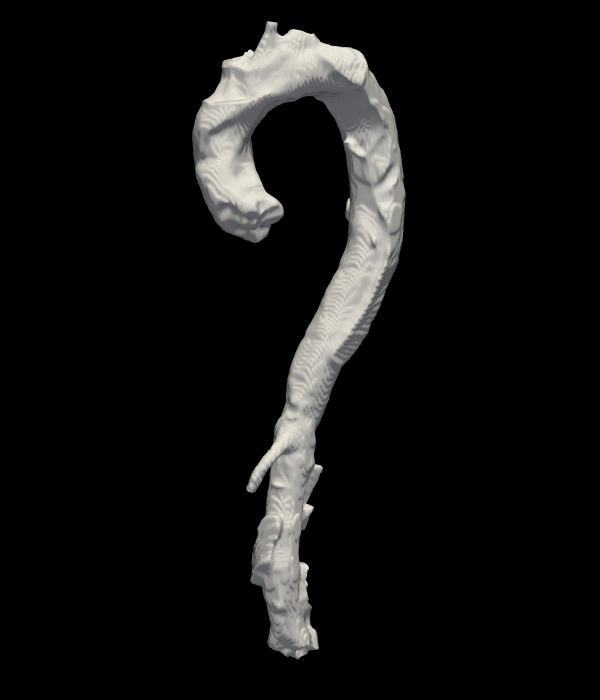}
    \caption{Patient 2, isoline $0.2$.}
\end{subfigure}
\begin{subfigure}[b]{0.3\textwidth}
    \centering
    \includegraphics[width=1.0\linewidth]{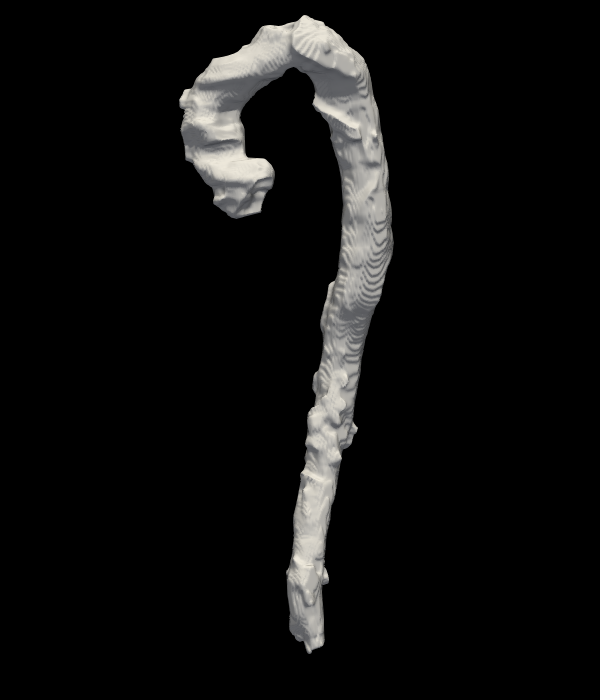}
    \caption{Patient 3, isoline $0.3$.}
\end{subfigure}

\vspace{0.5 cm}

\begin{subfigure}[b]{0.3\textwidth}
    \centering
    \includegraphics[width=1.0\linewidth]{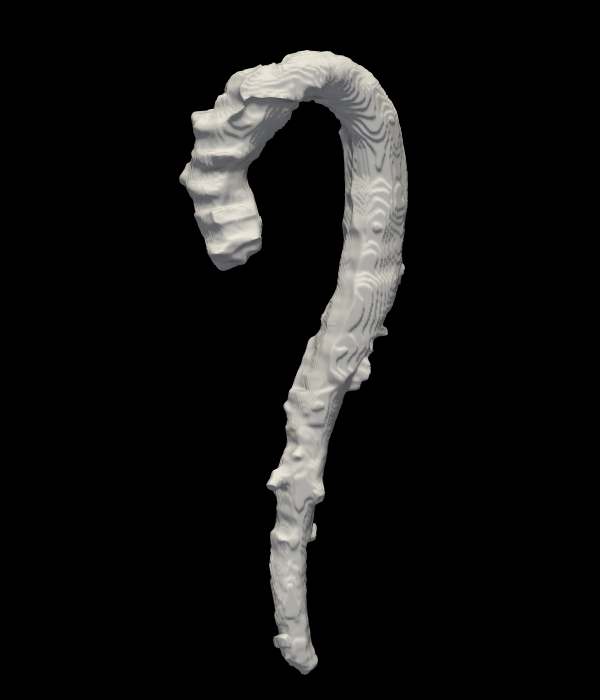}
    \caption{Patient 4, isoline $0.3$.}
\end{subfigure}
\begin{subfigure}[b]{0.3\textwidth}
    \centering
    \includegraphics[width=1.0\linewidth]{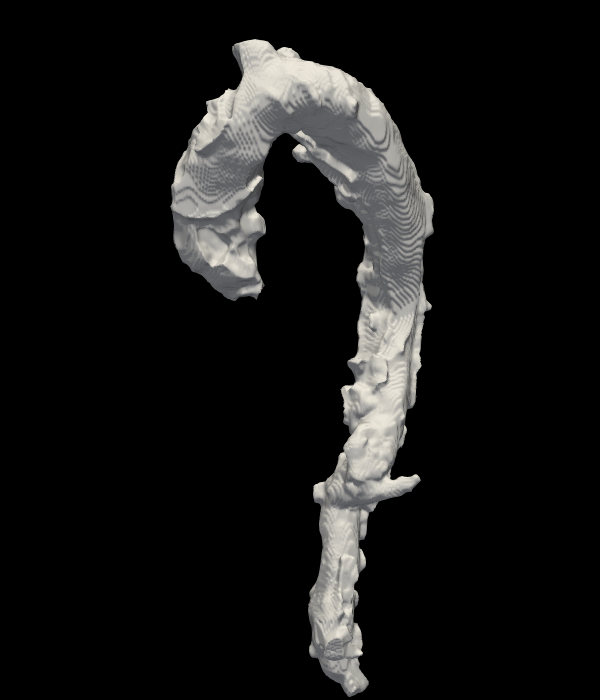}
    \caption{Patient 5, isoline $0.17$.}
\end{subfigure}
\begin{subfigure}[b]{0.3\textwidth}
    \centering
    \includegraphics[width=1.0\linewidth]{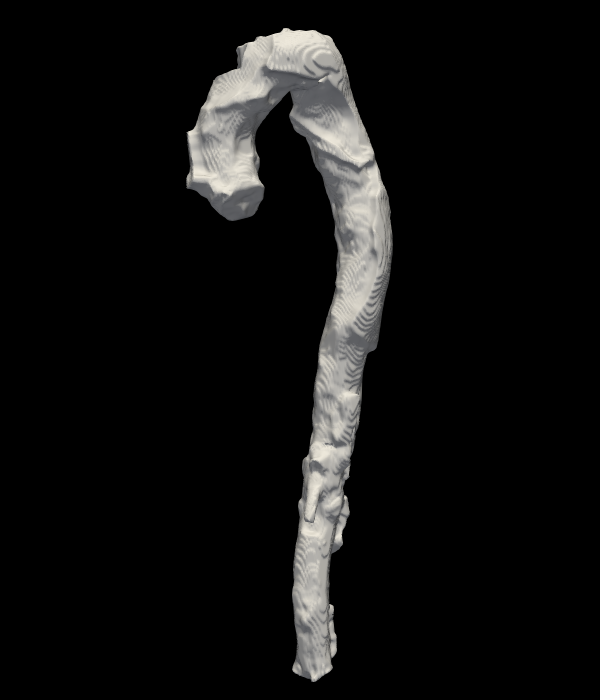}
    \caption{Patient 6, isoline $0.3$.}
\end{subfigure}

\caption{3D rendering of the segmentation results for the data sets.} 
\label{Result_g_subsurf_3D}
\end{figure}

Fig.\ref{Result_g_subsurf_3D} shows the results of the GSUBSURF segmentation in a 3D view for the dataset. The segmentation is stopped when the $L_2$ norm between the previous and current solution is lower than a given tolerance $\text{tol} = 0.001$. The grid size for each patient is taken equal to the voxel size (see Table \ref{spacing_patients_data}). The edge detector parameter is $K = 100000$ for all the patients, $\tau = 0.5$, $\omega_{ad} = 1.0$, $\omega_{dif} = 0.05$ and $\varepsilon = 0.001$.

\bigskip

We estimate the accuracy of our segmentation results using the Hausdorff distance \cite{Huttenlocher_1993}. Given two finite sets of points, $A = \{ a_1, \cdots, a_m\}$ and $B = \{b_1, \cdots, b_l \}$, the mean Hausdorff distance is defined as
\begin{equation}
    MHD(A,B) = max (mhd(A, B), mhd(B,A))
\end{equation}
where:
\begin{equation}
    mhd(A,B) = \frac{1}{m} \sum_{i = 1}^{m} \underset{b \in B}{min}~||a_i - b||
\end{equation}
and $|| \cdot ||$ is the Euclidean distance between points of $A$ and $B$. The set $A$ consists of voxels (points) corresponding to the reference segmentation (manual segmentation) of the aorta, and $B$ represents the set of the GSUBSURF segmentation results. We performed the manual segmentations using the software \textit{3D Slicer}.

\begin{table}[H]
\centering
\caption{Mean Hausdorff distance of segmentation results}

    \begin{tabular}{|c|c|c|c|c|c|c|}
    \hline
     Patients    & 1     & 2      & 3      & 4       & 5       & 6\\
     \hline
     MHD         & 1.85  & 2.2    & 1.57   & 1.72    & 2.78    & 1.9   \\
     \hline
    \end{tabular}
    \label{tab_seg}
\end{table}

Table \ref{tab_seg}. shows that the segmentation results and the reference segmentations overlap, with the maximum mean Hausdorff distance equal to $2.78$ for patient 5, which represents the worst-case scenario. That value corresponds to $3$ voxels in $x$ or $y$ directions but less than $2$ voxels in $z$ direction. These results suggest that our segmentation framework achieved a high level of overlap with the reference segmentations, indicating good performance. Nevertheless, as non-expert annotations generate the reference segmentation (manual), these results should be interpreted as indicative rather than definitive.

\newpage
\section{Application to the diagnosis of vasculitis}
Large-vessel vasculitis (LVV) is the most common type of vasculitis affecting predominantly the aorta and has two major variants: Giant Cell Arteritis (GCA) and Takayasu Arteritis (TA) \cite{Slart_2018, Dejaco_2018, Jennette_2013, Stenova_2009}. TA usually affects people before the age of 50, while GCA affects people older than 50 \cite{Jennette_2013}. Various approaches are used to diagnose LVV, including biopsy and imaging with FDG-PET/CT. Imaging by FDG-PET/CT is a non-invasive approach that can detect the disease at its early stage \cite{Arnaud_2009}.

\subsection{Diagnosis of LVV by FDG-PET/CT imaging}

The literature review has revealed that most available works on the diagnosis of LVV using FGD-PET/CT imaging have been conducted by experts in nuclear medicine, radiology, rheumatology, and vascular surgery. Therefore, our work is not to presume to replace them, but to offer a mathematical/software solution that can help them in their work. There is consensus about several questions, including recommended procedures for the diagnosis of LVV. For instance, a high FDG uptake in the wall of the aorta may efficiently guide the diagnosis of LVV\cite{Balink_2014}. In clinical routine, an experienced nuclear medicine physician performs a visual analysis, which is considered positive when the FDG concentration, which we call uptake, in the region of interest (the inflamed aortic regions) is higher than that in the reference tissue (the liver) \cite{Arnaud_2009}. In the authors' recommendations papers \cite{Dejaco_2018, Dejaco_2024}, for instance, the assessment is made through visual analysis of the affected aorta region uptake, and comparison is made with the liver uptake as background, when the visual analysis of the vessel is "unclear". In another recommendation paper \cite{Slart_2018}, the authors propose the use of a standardized 0-to-3 visual grading scale as follows:
\begin{itemize}
    \item \textbf{0} = no uptake
    \item \textbf{1} = low-grade uptake (vascular uptake $<$ liver uptake)
    \item \textbf{2} = intermediate-grade uptake (vascular uptake $\approx$ liver uptake)
    \item \textbf{3} = high-grade uptake (vascular uptake $>$ liver uptake)
\end{itemize}
where grade 2 is "possibly positive", and grade 3 is considered positive for active LVV. The authors of \cite{Slart_2018} emphasize that the visual analysis is not always enough for an optimal assessment. Therefore, they discuss some (semi-) quantitative methods for further classification. The target-to-background ratio (TBR) approach uses various structures, such as the lung, liver, or blood pool, as a background. The SUV represents the FDG uptake within the patient's organs, normalized by the injected dose of FDG \cite{Boellaard_2009}. The use of the liver as background is the most used approach \cite{Slart_2018} when interpreting FDG-PET images. In practice, ROIs are manually drawn around the target aorta regions, and parameters such as SUV$_{max}$ and SUV$_{mean}$ are calculated within the ROIs. The authors of \cite{Lmfeld_2018} recommend defining a spherical volume that encompasses the aorta wall in the aorta segment with the highest FDG uptake. For the background organ, a sphere of 3 cm diameter is used to define the ROI in the right lobe of the liver to minimize the chance of including veins and arteries that run through the liver \cite{Slart_2018}. As the authors of \cite{Boellaard_2009} emphasize in their work, the parameters SUV$_{max}$ and SUV$_{mean}$ depend strongly on the definition of the ROI. Therefore, there is a real need to define precise ROIs to make the quantitative analysis robust and reproducible. Several works have addressed these questions before us. The authors of \cite{Hatt_2007} presented a literature review of the proposed segmentation approaches. We can read there that the proposed methods include thresholding and region growing, and are directly applied to the PET image data. In addition, the work \cite{Hatt_2007} was not dedicated to vasculitis, but rather to the interpretation of PET in general, which is also used for cancer diagnosis. These works suffer from poor contrast in PET and recurrent noise. Therefore, our approach, which involves segmenting the ROIs on CT with better resolution than PET, followed by alignment with PET, yields significant improvements in the definition of ROIs.

\subsection{Application}

In this section, we present the application of the aorta centerline extraction and segmentation to the interpretation of PET images. The data are obtained for patients suspected of having vasculitis.

\subsubsection{Definition of the regions of interest}

One of our objectives in this work is to define precise ROIs in the liver and the aorta to assess vasculitis. To obtain the liver ROI, we segment the liver from the CT data (see \cite{Konan_2024}), then find its centroid, and define a spherical volume with a diameter of 3 cm. For the aorta, contrary to what is done in \cite{Lmfeld_2018}, where the ROI is defined in the most affected aorta region, we define multiple ROIs using the centered path points and the aorta segmentation. Each centered path point is assigned the distance from the segmented aorta boundaries plus one PET voxel size (see Table \ref{spacing_patients_data_pet}) to ensure that a sphere centered at that point (voxel) encompasses the aorta edges. Therefore, for each centered path point, we define an aorta ROI where we compute the maximum SUV, i.e., SUV$_{max}$. 

\newpage
\subsubsection{Results}

In this section, we present the results of our quantitative analysis of FDG-PET/CT image data to assess vasculitis. The CT and FDG-PET images are aligned (registered) to find the FDG-PET image regions corresponding to the extracted ROIs from the CT image. In this work, we assume that the FDG-PET image voxel values correspond to the SUV in each voxel. 

\begin{figure}[H]
\centering
\includegraphics[width=0.45\linewidth]{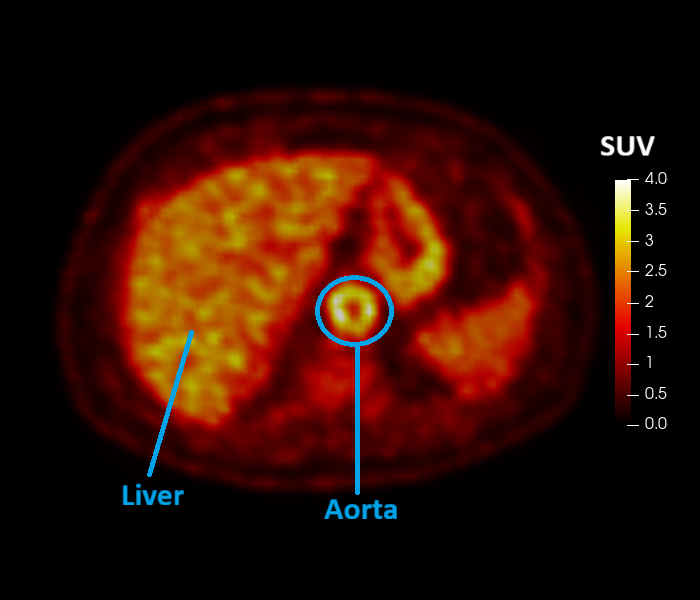}
\includegraphics[width=0.45\linewidth]{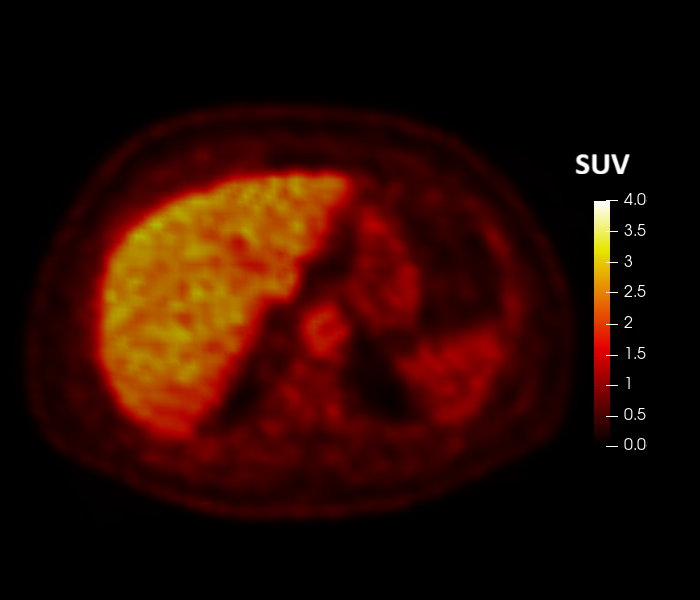}
\captionsetup{skip=10pt}
\caption{PET slice view of the aorta's highest SUV region before treatment (left) and the corresponding slice after treatment (right).} 
\label{pet_slices_before_after}
\end{figure}

Fig.\ref{pet_slices_before_after} shows the FDG uptake in the same plane for a patient before and after treatment. We detected this slice by analysing the SUV ratios 
\begin{equation}
    ratio = \frac{SUV_{max~aorta}}{SUV_{mean~liver}} \label{formula_Ratio}
\end{equation}
along the aorta, and it corresponds to the maximal value. Before treatment, a visual analysis suggests a grade 2 (aorta uptake $>$ liver uptake), whereas the case after treatment is not trivial according to the visual analysis. This example demonstrates that an automatic analysis, as we suggest, can detect aorta regions with high FDG uptake and properly define the grade (see Fig.\ref{pet_slices_before_after} and discussion in the next paragraph). This means that our framework is capable of providing reproducible parameters (SUV ratios) for the diagnosis of LVV using FDG-PET/CT image data.

\bigskip

In the following example, we present FDG-PET image interpretation based on the SUV ratios computed along the aorta.

\begin{figure}[H]
\centering
\includegraphics[width=0.45\linewidth]{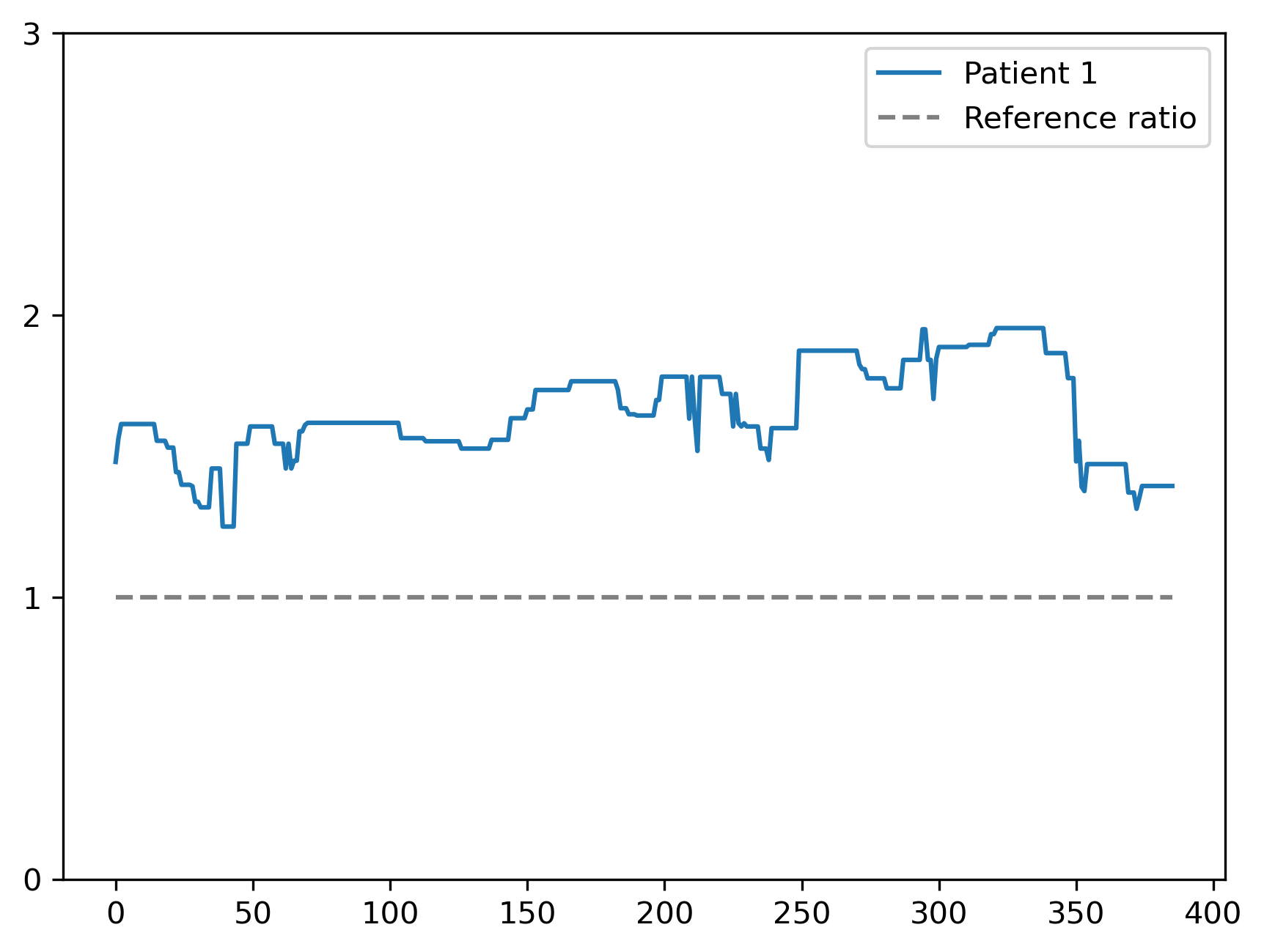}
\includegraphics[width=0.45\linewidth]{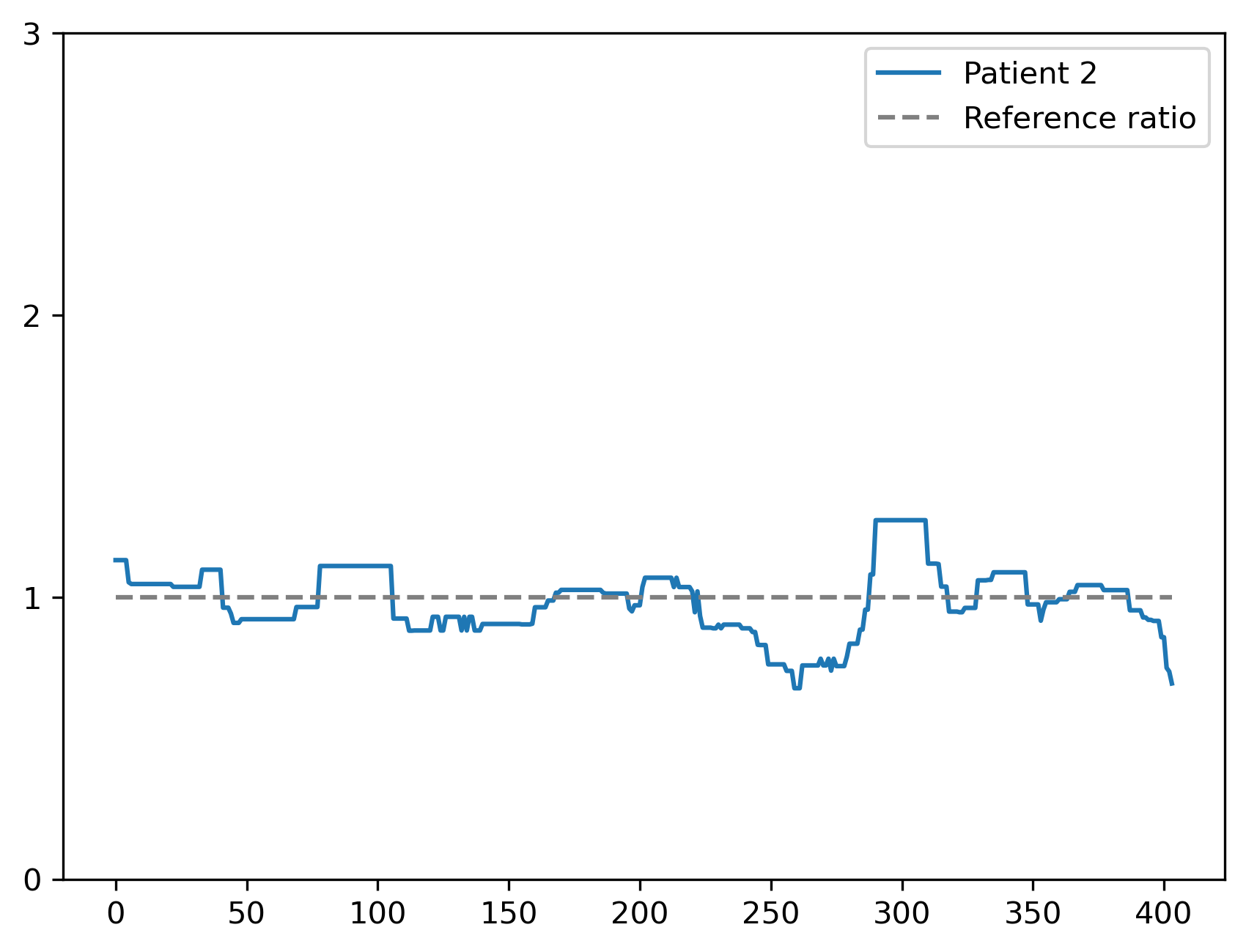}
\includegraphics[width=0.45\linewidth]{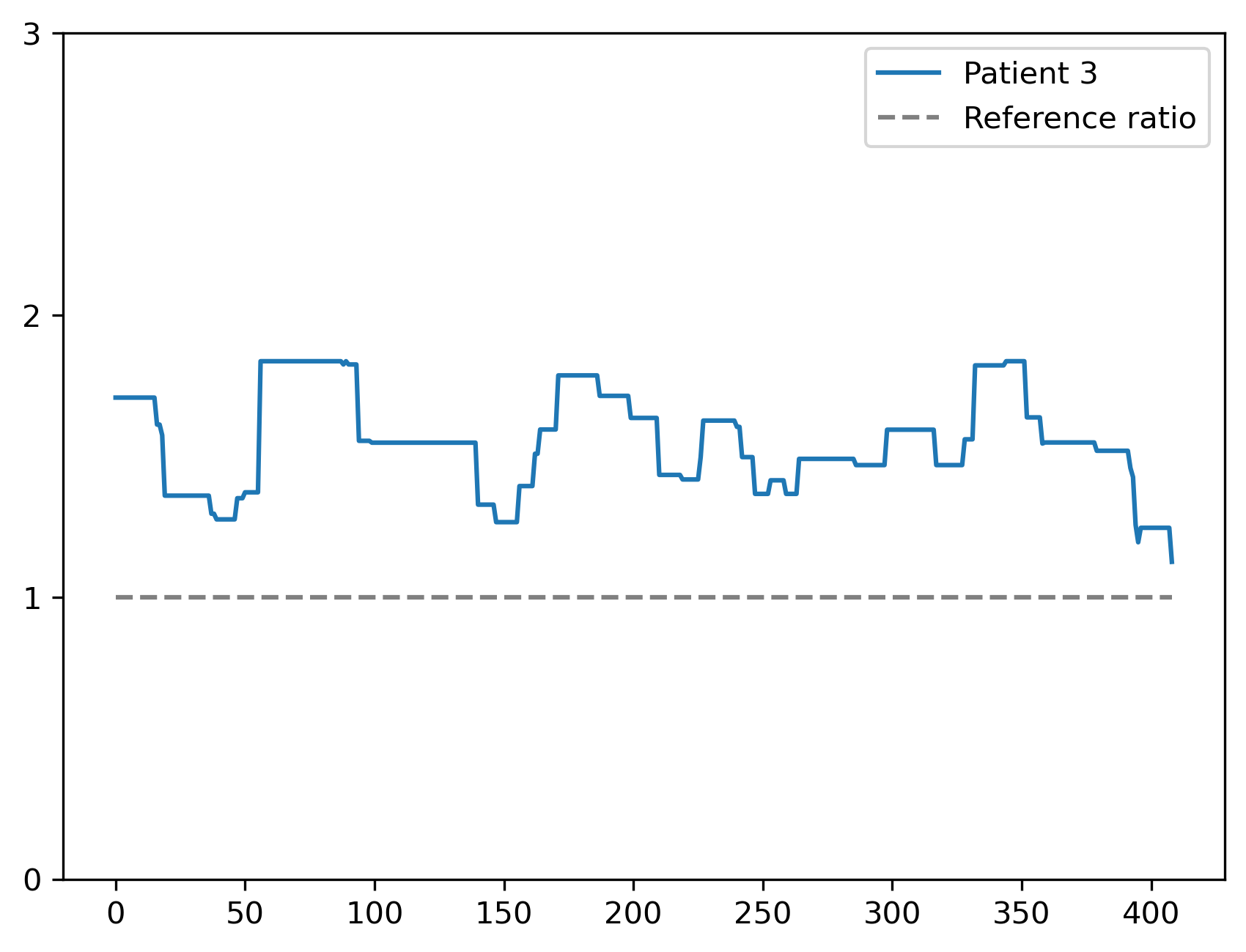}
\includegraphics[width=0.45\linewidth]{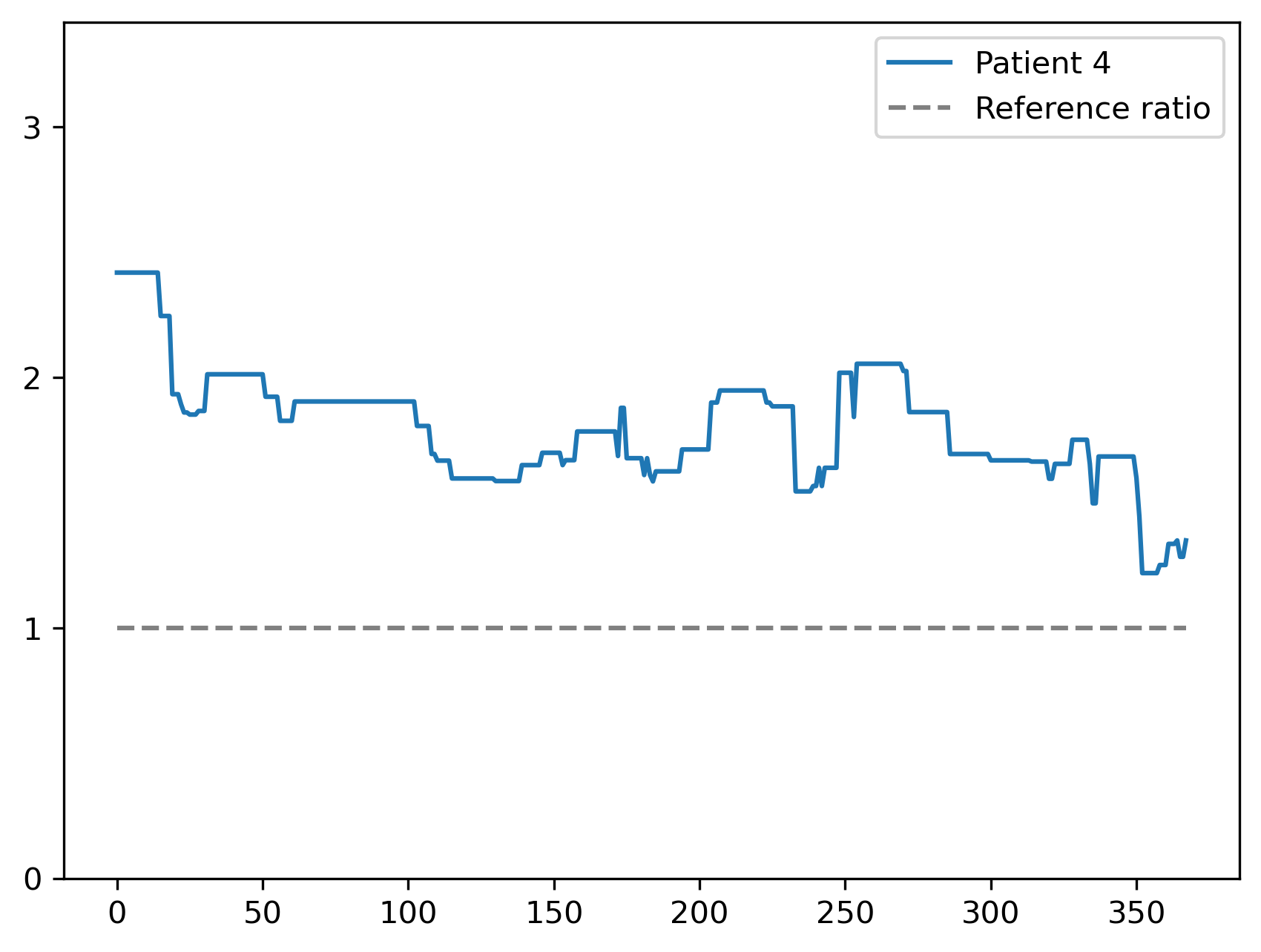}
\includegraphics[width=0.45\linewidth]{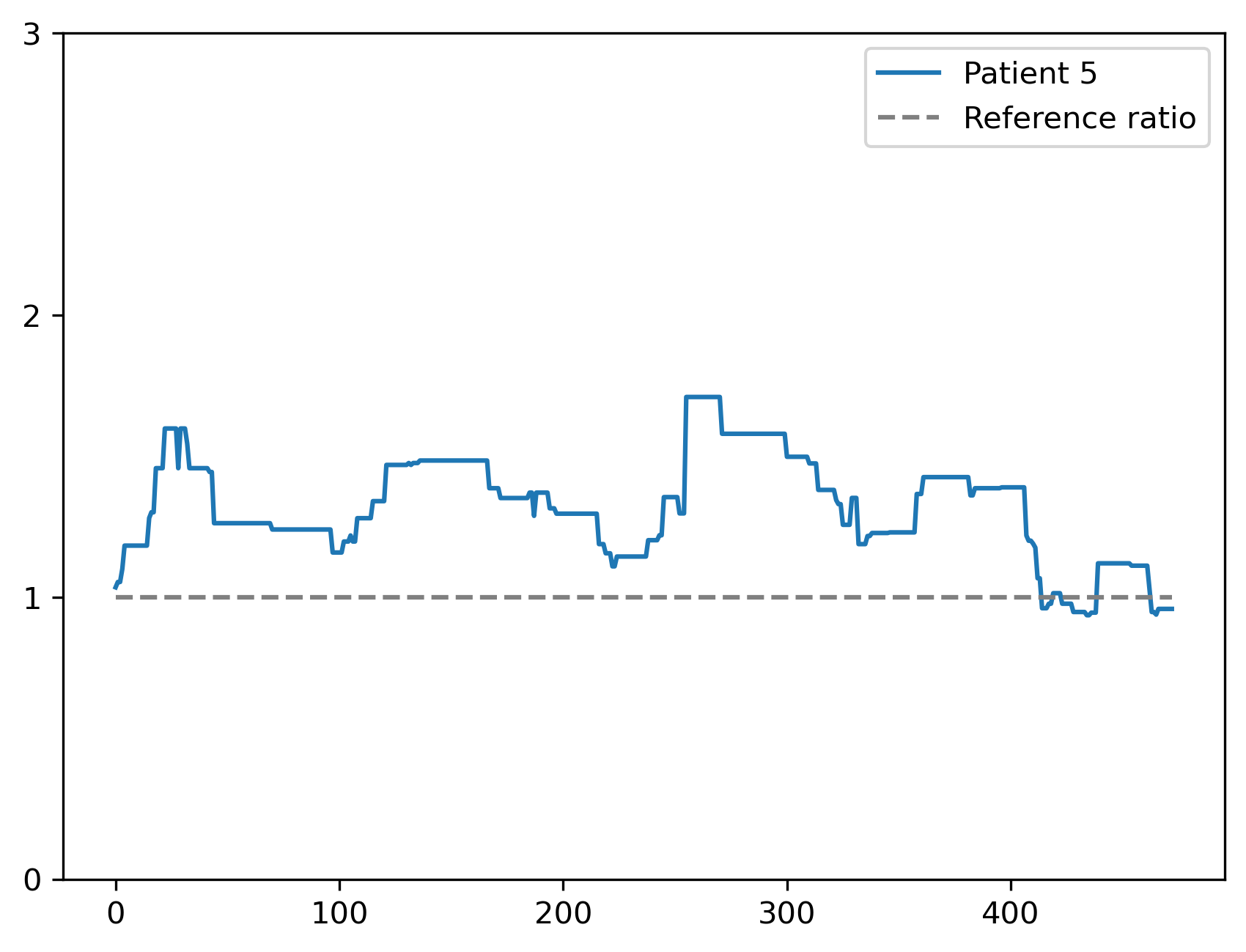}
\includegraphics[width=0.45\linewidth]{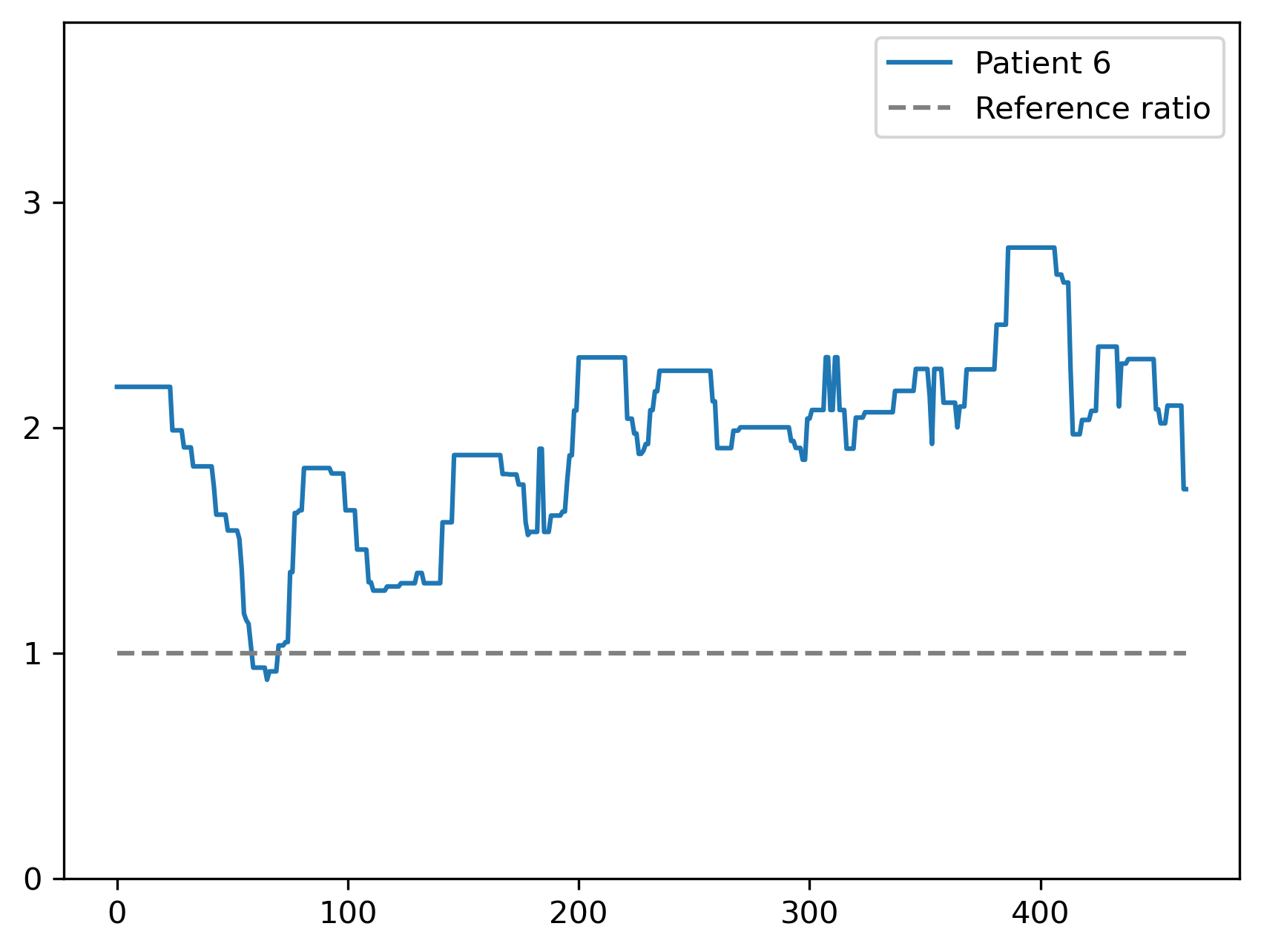}
\caption{Ratio vascular uptake and liver uptake along the centered path for the dataset.} 
\label{ratio_aorta_segment_liver}
\end{figure}

Fig.\ref {ratio_aorta_segment_liver} presents the calculated ratios along the aorta for our dataset. The x-axis corresponds to the number of points in each aorta centerline. The y-axis corresponds to the calculated ratio (\ref{formula_Ratio}). The gray line (dashed) corresponds to a ratio equal $1$. The graphs for patients 1, 3, and 4 indicate that vascular uptake significantly exceeds liver uptake along the aorta. This result suggests that the inflammation is very extended and a treatment is required. The graphs for patients 5 and 6 indicate that the inflammation is located in specific regions of the aorta. Since we have the aorta centerline, we can precisely identify these regions using the coordinates of the path points. The graph for patient 1 shows that the aorta uptake is greater than the liver uptake along the aorta. In contrast, the graph for patient 2 shows smaller ratios compared to patient 1, as well as some regions where liver uptake is greater. Since these two image data were provided from the same patient, before and after treatment, we can conclude that the treatment is having a positive effect on the aorta inflammation. The visual analysis was not straightforward in assessing whether the vascular uptake was higher than the liver uptake in the right image of Fig.\ref{pet_slices_before_after}. However, with our proposed quantitative approach, we can surely conclude that for that slice, the liver uptake is larger, meaning the treatment was successful. This example demonstrates that quantitative analysis can aid practitioners in their assessment when visual analysis alone is insufficient for concluding. 

\section{Conclusion}
In this paper, we proposed a framework for segmenting the aorta from non-enhanced CT data. The segmented aorta is used to define precise ROIs for the diagnosis of aortic vessel vasculitis using FDG-PET/CT image data. The three steps of our proposed approach are robust enough to handle the variability of the aorta shape. We proposed a new potential function to extract a path within the aorta using the minimal path approach with keypoint detection, given two endpoints. We utilize a 3D curve evolution model to move the initial path to the centerline of the aorta. The original work \cite{Mikula_Urban_2014} computes the velocity vector field from available segmentation. However, we do not have the segmented aorta as input. Therefore, we use edge detection and threshold to extract the aorta edges to construct the velocity vector field. This application demonstrates that the model proposed in \cite{Mikula_Urban_2014} is robust enough to perform well even when some edges are missing. In the final step, we apply the GSUBSURF method, using the aorta centerline to construct the initial condition, to extract the aorta surface. Our proposed framework is capable of processing a given FDG-PET/CT image data with less user interaction. The framework can assess vasculitis when the visual analysis is not straightforward. Our future work will be dedicated to the automatic detection of endpoints for minimal path extraction, aiming to make the framework fully automatic. 

\bigskip

\noindent\textbf{Aknowledgments.}
\noindent We thank TatraMed Software s.r.o. for technical support and to INLANET for funding. This project has received funding from the European Union’s Horizon 2020 research and innovation programme under the Marie Skłodowska-Curie grant agreement No 955576.

\newpage
\bibliographystyle{plain}
\bibliography{tmmp000}
\end{document}